\documentclass[prl,twocolumn,preprintnumbers,amsmath,amssymb,superscriptaddress,longbibliography]{revtex4-2}

\usepackage{color}
\usepackage{graphicx}
\usepackage{dcolumn}
\usepackage{bm}
\usepackage{hyperref}
\usepackage{enumitem}
\usepackage{balance}
\usepackage{comment}
\usepackage{cleveref}

\usepackage{amsthm}

\begin{document}
\title{A magic monotone for faithful detection of non-stabilizerness in mixed states}

\author{Krzysztof Warmuz}
\affiliation{New York University Shanghai, NYU-ECNU Institute of Physics at NYU Shanghai, 567 West Yangsi Road, Shanghai, 200124, China.}

\author{Ernest Dokudowiec}
\affiliation{Gonville \& Caius College, University of Cambridge, Cambridge, CB2 1TA, United Kingdom}
\affiliation{New York University Shanghai, NYU-ECNU Institute of Physics at NYU Shanghai, 567 West Yangsi Road, Shanghai, 200124, China.}

\author{Chandrashekar Radhakrishnan}
\affiliation{Department of Computer Science and Engineering,  NYU Shanghai, 567 West Yangsi Road, Shanghai, 200124, China.}

\author{Tim Byrnes}
 \email{tim.byrnes@nyu.edu}
\affiliation{New York University Shanghai, NYU-ECNU Institute of Physics at NYU Shanghai, 567 West Yangsi Road, Shanghai, 200124, China.}
\affiliation{State Key Laboratory of Precision Spectroscopy, School of Physical and Material Sciences, East China Normal University, Shanghai 200062, China}
\affiliation{Center for Quantum and Topological Systems (CQTS), NYUAD Research Institute, New York University Abu Dhabi, UAE.}
\affiliation{Department of Physics, New York University, New York, NY 10003, USA}

\begin{abstract}
We introduce a monotone to quantify the amount of non-stabilizerness (or magic for short), in an arbitrary quantum state.  The monotone gives a necessary and sufficient criterion for detecting the presence of magic for both pure and mixed states. The monotone is based on determining the boundaries of the stabilizer polytope in the space of Pauli string expectation values.  The boundaries can be described by a set of hyperplane inequations, where violation of any one of these gives a necessary and sufficient condition for magic.  The monotone is constructed by finding the hyperplane with the maximum violation and is a type of Minkowski functional.  We also introduce a witness based on similar methods.  
The approach is more computationally efficient than existing faithful mixed state monotones such as robustness of magic due to the smaller number and discrete nature of the parameters to be optimized. 
\end{abstract}

\date{\today}

\maketitle

\paragraph{Introduction}
The Gottesmann-Knill theorem 
\cite{gottesman1998heisenberg,gottesman1998talk} is one of the seminal results in the field of quantum computation, which states that any quantum circuit that only consists of Clifford gates can be simulated on a classical computer in polynomial time \cite{aaronson2004improved,anders2006fast}.  The reason for this remarkable 
result is that such quantum circuits, called stabilizer or Clifford circuits, have a special symmetry, where the output of the circuit can be only one of an 
enumerable number of stabilizer states.  Stabilizer states are simultaneous eigenstates of Pauli strings, and using the fact that under Clifford 
transformations such Pauli strings transform into other Pauli strings, one may efficiently keep track of the evolution in the quantum circuit \cite{gottesman1997stabilizer,nielsen2002quantum}.  Equivalently, in 
the Heisenberg picture, the operator evolution greatly simplifies due to the lack of operator growth thanks to the nature of Clifford transformations 
\cite{aharonov2023polynomial,ermakov2024unified}. Since such Clifford circuits can be efficiently evaluated on a quantum computer, it follows that for a quantum computer to perform 
a task that is intractable for a classical computer, it must be capable of non-Clifford operations or have non-stabilizer states available to it.  
Such non-stabilizer states and operations, also called magic states and gates, can be considered a resource to perform universal quantum computation \cite{veitch2014resource,veitch2012negative,mari2012positive,howard2017application}. 

A natural task in this context is then to detect and quantify the amount of non-stabilizerness (or magic) in a given quantum state. Restricting our 
discussion only to qubit systems (as opposed to qudits), one of the best known definition is the robustness of magic (RoM), which gives a faithful criterion for the detection of magic for both pure and mixed states, and satisfies several properties that make it a monotone
\cite{vidal1999robustness, howard2017application}.  Several alternative faithful monotones were proposed by Seddon, Regula, Campbell and co-workers improving the compatibility at the expense of introducing a failiure 
probability \cite{seddon2021quantifying}.  The main drawback of these methods are that in an exact calculation they are highly numerically intensive since 
they involve the number of stabilizer states, which grow superexponentially with the number of qubits. Other quantities tend to be easier to calculate but 
have other drawbacks.  The stablizer extent \cite{bravyi2016improved,bravyi2016trading,bravyi2019simulation}, stabilizer nullity, dyadic monotone 
\cite{beverland2020lower}, stabilizer R{\'e}nyi entropy \cite{leone2022renyientropy,leone2024stabilizerentropies,haug2023stabilizerentropies,oliviero2022measuring}, GKP magic \cite{hahn2022quantifying}
and Bell magic \cite{haug2023scalablemeasures} are only 
valid for pure states. Sum negativity, mana and related measures like Thauma 
\cite{veitch2012negativequasiprobability,veitch2014resource,wang2020thauma, wang2019quantifyingmagic, koukoulekidis2022wignernegativity}
and have been successfully computed using Monte Carlo methods \cite{pashayan2015quasiprobabilities, kulikov2024minimizingnegativity}  but do not work for qubit systems.
The stabilizer norm can be applied to both pure and mixed states, and gives a sufficient criterion for magic, however, does not give a very sensitive criterion in many cases \cite{campbell2011catalysis,howard2017application}. It is therefore desirable to obtain a quantifier for magic that is more easily 
computable, works for mixed states, and satisfies key properties such as faithfulness.

In this paper, we introduce a new monotone to detect and quantify the amount of magic in a given state.  Our approach is based upon determining the boundaries of the stabilizer polytope, which is the set of states that can be formed by a probabilistic combination of stabilizer states.  By giving an explicit criterion for the facet hyperplanes in the space of Pauli string expectation values, we give necessary and sufficient conditions for a magic state.  This can be formed into a monotone which quantifies the amount of magic.  We also introduce a witness which is convenient for numerical computation and show their effectiveness in detecting magic for mixed states.

\paragraph{Stabilizer states and magic}
Consider an $ N $-qubit system and denote the pure stabilizer states as $ |S_i \rangle $, which are simultaneous eigenstates of $ 2^N $ commuting Pauli strings taking the form $ P_k = \bigotimes_{n=1}^N P_n^{(l)} $, where $ P_n^{(l)} \in \{ I_n, X_n, Y_n, Z_n \} $ are Pauli matrices on site $ n $ with $ l \in [0,3] $.   We order the Pauli strings according to the digits of $ k \in [0, D^2-1] $ in base 4, such that $ P_0 = I^{\otimes N} $, where $ D = 2^N $ is the Hilbert space dimension. There are a total of $ D_S = 2^N \prod_{n=1}^N (2^n +1) \sim 2^{N^2/2} $ pure stabilizer states, so that the label runs from $ i \in [1, D_S] $ \cite{gross2006hudson}.  More generally, stabilizer states can be formed by a probabilistic mixture of pure stabilizer states
\begin{align}
\rho_S = \sum_{i=1}^{D_S} p_i | S_i \rangle \langle S_i | ,
\label{mixedstabilizer}
\end{align}
where $ 0 \le p_i \le 1 $ are probabilities with $ \sum_{i=1}^{D_S} p_i = 1$. The set of stabilizer states is known as the stabilizer polytope and consists of the convex hull of the pure stabilizer states.

A non-stabilizer state can be defined as any state that cannot be written in the form (\ref{mixedstabilizer}). By allowing $ p_i $ to take negative values, it becomes possible to write any arbitrary state $ \rho $ as an affine mixture of pure stabilizer states.  Using this, a suitable quantifier for the magic of a general state is the robustness of magic (RoM), defined as
\begin{align}
R(\rho) = \min \left\{ 
\sum_{i=1}^{D_S} |x_i | : \rho =  \sum_{i=1}^{D_S} x_i | S_i \rangle \langle S_i |  \right\} .
\end{align}
Here, the minimization is performed over the real parameters $ x_i $, which may be negative in this case. A necessary and sufficient criterion for presence of magic is $ R(\rho) > 1 $, and all stabilizer mixtures (\ref{mixedstabilizer}) have $ R(\rho) = 1$. Due to the superexponential number of such parameters typically this is a highly intensive numerical problem such that the largest system that can be calculated is $ N \sim 5 $ \cite{howard2017application}. 

Another witness for magic is the stabilizer norm, defined as \cite{campbell2011catalysis,howard2017application}
\begin{align}
|| \rho ||_{\text st} = \frac{1}{2^N} \sum_{k=0}^{D^2-1} | \langle P_k  \rangle | ,
\label{stnorm}
\end{align}
where $ \langle P_k \rangle = \text{Tr} (\rho P_k ) $, and detects magic when $ || \rho ||_{\text st}  > 1 $.  For $ N = 1$, this is a necessary and sufficient criterion for magic and recovers the well-known octahedral stabilizer polytope which gives the boundary between magic and stabilizer states:
\begin{align}
| \langle X \rangle | + | \langle Y \rangle | +| \langle Z \rangle |  = 1 .
\label{octahedron}
\end{align}
For $ N \ge 2 $, the stabilizer norm is however only a sufficient condition, and some magic states are missed.

\paragraph{Polytope boundaries}

We now formulate a general method to find stabilizer polytope boundaries, with the aim of generalizing the result (\ref{octahedron}) to arbitrary $ N $.  First let us discuss the space which the polytope exists in.  We shall work in the space $ \cal P $ defined by the expectation values of Pauli strings, such that any state $ \rho $ is represented by a vector of length $ D^2 $
\begin{align}
   \langle \vec{P} \rangle = ( \langle P_0 \rangle, \langle P_1 \rangle, \dots ,\langle P_{D^2 - 1} \rangle ) ,
\end{align}
where $ \vec{P} $ denotes a vector formed by all the Pauli string operators (with $ +1 $ coefficients). The vector $ \langle \vec{P} \rangle $ contains full information of the density matrix $ \rho $ and naturally generalizes the space which the Bloch sphere exists in for $ N = 1$.  

The pure stabilizer states in ${\cal P}$-space, defined as $ \vec{S}_i = \langle S_i | \vec{P} | S_i \rangle $, take a characteristic form of having $ D $ non-zero elements each taking a value of $ \pm 1 $ and the remaining being zero (see Appendix).  The non-zero elements correspond to $ D $ mutually commuting Pauli strings, including the identity.   A general mixed stabilizer state (\ref{mixedstabilizer}) in ${\cal P}$-space then forms a convex polytope parameterized by the region
\begin{align}
\langle \vec{P} \rangle = \sum_{i=1}^{D_S} p_i \vec{S}_i ,
\label{polytopevec}
\end{align}
where $ 0 \le p_i \le 1 $.  By extending $ p_i $ to negative values it is possible to write an arbitrary state as an affine mixture in the same way as done with RoM. 

The boundary of the stabilizer polytope are formed by hyperplanes \cite{grunbaum1967convex} that pass through the stabilizer vectors $ \vec{S}_i $, given by the general form
\begin{align}
    \vec{a} \cdot \langle \vec{P} \rangle = b ,
    \label{hyperplane}
\end{align}
where $ \vec{a} $ is a $ D^2 $ dimensional vector and $ b $ is a constant. The polytope boundaries must take a linear form (as opposed to, for instance, a curved surface), due to the linear mixture of the stabilizers along a boundary.  Consider a particular face of the polytope consisting of a mixture of $ D_F $ stabilizers $ F = \{|S_1^F \rangle , \dots , | S_{D_F}^F \rangle  \} $, which are a subset of the all the stabilizers.  Along a polytope boundary, we have the mixture
\begin{align}
\rho_F = \sum_{j=1}^{D_F} p_j |S_{j}^F \rangle \langle S_{j}^F | .
\label{polytopeboundaryrho}
\end{align}
Here $ D_F < D_S $ such that some of the coefficients $ p_i $ have a zero value, giving the opportunity for them to turn negative.  In $ \cal P$-space, this appears as  $ \langle \vec{P} \rangle = \sum_{j=1}^{D_F} p_j  \vec{S}_{j}^F $, which is a parameterized form of a hyperplane, equivalent to (\ref{hyperplane}).  

\begin{figure}[t]
\includegraphics[width=\linewidth]{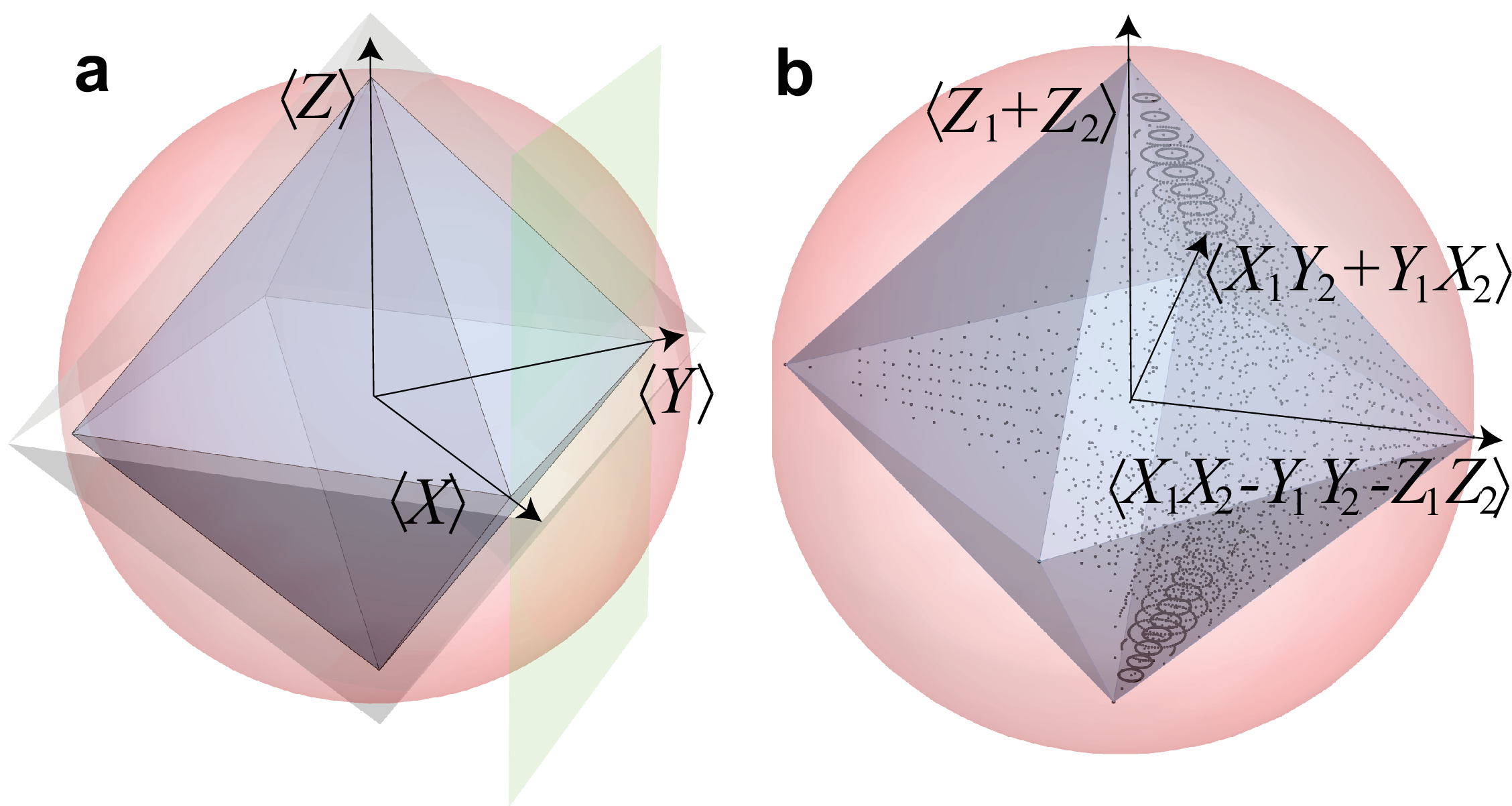}
\caption{Stabilizer polytope boundaries (\ref{polytopeboundary}) according to various choices of $ \vec{a} $.  (a) The polytope boundaries for $ N =1 $ corresponding to $ \pm \langle X \rangle \pm' \langle Y \rangle \pm \pm' \langle Z \rangle \le  1 $, where $ \pm, \pm' $ can be chosen independently (inner octahedron). Also shown is the polytope boundary $  \langle X \rangle + \langle Y \rangle \le 1 $ (vertical plane). The surface of the outer octahedron defines planes of constant $ {\cal M}(\rho) = 1.2 $. The Bloch sphere showing the boundary of all states is also shown.  (b) The polytope boundaries for $ N = 2 $ corresponding to $ -3 \le \langle X_1 X_2 - Y_1 Y_2 - Z_1 Z_2 \rangle \pm \langle X_1 Y_2 + Y_1 X_2 \rangle \pm' \langle Z_1 +Z_2 \rangle \le 1 $.  The dots correspond to points with $ {\cal W}(\rho)  = 0 $ for the Werner state $ \rho = (1-\mu) I/D + \mu |\psi \rangle \langle \psi | $ with $ |\psi \rangle = \cos \tfrac{\theta}{2} |00 \rangle +  e^{i \phi} \sin \tfrac{\theta}{2} |11 \rangle $. \label{fig1}  }
\end{figure}

The coefficients of the hyperplane must satisfy certain conditions in order that they form a valid boundary of the stabilizer polytope. We define a polytope boundary as any hyperplane that contains at least one point from the stabilizer polytope and defines the half-plane such that the polytope is on one side (see Fig. \ref{fig1}(a)).  Suppose we are given a particular $ \vec{a} $ which defines the slope of the hyperplane.  Then if we take 
\begin{align}
b(\vec{a}) \equiv \max_{i\in[1,D_S] } \vec{a} \cdot \vec{S}_i 
\label{bcond}
\end{align}
then this ensures that all mixed stabilizer states satisfy
\begin{align}
    \vec{a} \cdot  \langle \vec{P} \rangle \le b(\vec{a}) .
    \label{polytopeboundary}
\end{align}
Another bound can be obtained by replacing $\vec{a} \rightarrow - \vec{a} $, which corresponds to the lower bound of the polytope  (see Appendix). 

Now suppose we start with a candidate subset $ F $ of all the pure stabilizers which form a mixture of the form (\ref{polytopeboundaryrho}), which may or may not lie on a polytope boundary.  How do we determine whether $ F $ forms a polytope boundary?   First find the equation of the hyperplane that runs through all the stabilizers in $ F $ by demanding that (see Appendix)
\begin{align}
\vec{a} \cdot (\vec{S}_{j}^F - \vec{S}_{1}^F ) = 0 
\label{acond}
\end{align}
for all $ j \in [2, D_F ] $.  Depending upon the number of stabilizers $ M $ chosen, this may result in an underconstrained or overconstrained set of equations. In the overconstrained case, there may be no solution to (\ref{acond}) as no hyperplane exists to go through all the stabilizers in $ F $, meaning that $ F $ is not a polytope boundary.  In the underconstrained case, this will result in a set of hyperplanes with free parameters.  Once the coefficients that satisfy (\ref{acond}) are found, all $ \vec{a} \cdot \vec{S}_{j}^F $ equal a constant $ b $ for all $ j \in [1, D_F] $.  Then to see whether this is a polytope boundary, we must verify that it satisfies (\ref{polytopeboundary}), which can be equally written as
\begin{align}
    \vec{a} \cdot (\vec{S}_{1}^F -\vec{S}_i) \ge 0 .
    \label{polytopecond}
\end{align}
for all $ i \in [1,D_S] $. This condition demands that the hyperplane runs on one side of the polytope such that all stabilizer points are lower than it.  Thus if (\ref{acond}) and (\ref{polytopecond}) can be satisfied, we can conclude that $ F $ forms a polytope boundary.

\paragraph{Polytope boundary symmetries}

The stabilizer polytopes for multi-qubit states possess several symmetries due to the properties of stabilizer states \cite{heinrich2019robustness}.  First, due to the fact that Clifford unitaries map a pure stabilizer state onto another pure stabilizer state $ U_C |S_i \rangle \propto |S_{C(i)} \rangle$, a Clifford transformation of the states on the polytope boundary  (\ref{polytopeboundaryrho}) gives another polytope boundary $ U_C \rho_F U_C^\dagger $.  Then given that (\ref{polytopeboundary}) is a polytope boundary, then the same inequation with $ \vec{P} \rightarrow  U_C \vec{P} U_C^\dagger $ is also a polytope boundary, which is a permutation of the Pauli strings up to sign changes.  
Another symmetry is due to the spin flip symmetry of individual qubits.  Here we consider a spin flip to be along one of the stabilizer axes $ X, Y, Z $.  This consists of changing sign of one Pauli matrix on a site $ n $, i.e. $ P_n^{(l)} \rightarrow -P_n^{(l)}  $ for $ l \in [1,3] $.  In the Pauli vector $ \vec{P} $, this will change the signs of $ 4^{N-1} $ of the $ P_n $.  Then given that (\ref{polytopeboundary}) is a polytope boundary, the same inequation with this transformation is also a polytope boundary.  Multiple spin flips can be applied, in combination with Clifford transformations, which gives a family of hyperplanes which together define the boundary of the polytope (see Appendix). 

Another important simplification is that the hyperplane vector $ \vec{a} $ only takes integer components  $ a_k \in \mathbb{Z} $.  The reason for this originates from the fact that the stabilizer vectors only have components that are $ [\vec{S}_i]_k \in \{0, \pm 1 \} $, such that any hyperplane running through them must also take coefficients that are integral (see Appendix).  This also implies that $ b(\vec{a}) \in \mathbb{Z}  $.

\paragraph{Example polytope boundaries}

Figure \ref{fig1} shows some example polytope boundaries determined by the above procedure. Figure \ref{fig1}(a) shows the familiar single qubit case.  Choosing any three non-orthogonal stabilizers for $ F $ gives the 8 hyperplanes corresponding to the faces of the octahedral stabilizer polytope.  Choosing two  stabilizers (e.g. along the $ \langle X \rangle  $ and $ \langle Y \rangle $-axis) defines a boundary only along one edge of the polytope, but nevertheless it is a valid polytope boundary. For $ N = 2 $ (Fig. \ref{fig1}(b)), we use a similar procedure to construct a subset $ F $ that corresponds to a fully constrained problem. This is performed by starting with a seed set of stabilizers which are in the vicinity of a state of interest, and then continue to add stabilizers until (\ref{acond}) and (\ref{polytopecond}) are fully constrained.  For the example shown, we find that 33 stabilizers fully constrain the hyperplane, each giving a solution of $ \vec{a} $ with 7 Pauli strings with equal weight.  An example of eight such hyperplanes is shown in Fig. \ref{fig1}(b), which forms part of the polytope boundary in the 16 dimensional space, in agreement with Ref. \cite{reichardt2009quantum} obtained with alternative methods. It is noteworthy to add that for systems consisting of 2 or more qubits, the stabilizer polytope boundaries are given by several families of hyperplanes which are not related to one another by any of the Clifford symmetries. Plotting zero magic Werner states for the states $  \cos \tfrac{\theta}{2} |00 \rangle +  e^{i \phi} \sin \tfrac{\theta}{2} |11 \rangle $ we find that all fall within the hyperplanes.  However, in the negative $ \langle X_1 X_2 - Y_1 Y_2 -Z_1 Z_2 \rangle $ direction, there are other polytope boundaries (not plotted) which is the reason that Werner state does not reach the edge of the octahedron.

\paragraph{Necessary and sufficient conditions}

In deriving (\ref{acond}) and (\ref{polytopecond}), we took the approach of deriving the polytope boundary that passes through a given subset of stabilizers $ F $. In fact, it is not necessary to specify $ F $ to obtain a valid polytope boundary since given any $ \vec{a} $, the bound may be evaluated by 
(\ref{bcond}).  The stabilizer polytope is then defined by the set of points in $ \cal P $-space which satisfies
\begin{align}
SP_N = \{ \vec{a} \cdot \langle \vec{P} \rangle  \le b (\vec{a}) , \forall \vec{a} \in \mathbb{Z}^{D^2}   \}
\label{allpolytopeboundaries}
\end{align}
A violation of (\ref{allpolytopeboundaries}) is then a necessary and sufficient condition for the detection of magic.  This is a sufficient condition as already shown, since any stabilizer mixture must follow (\ref{polytopeboundary}).  It is also a necessary condition because no magic states can exist inside the stabilizer polytope (see Appendix).

\paragraph{Magic monotone}

Based on the above we can define the magic monotone
\begin{align}
{\cal M}(\rho)  = \max_{  \{ \vec{a} \in \mathbb{Z}^{D^2} , a_0 = 0     \}  }  [ \frac{ \vec{a} \cdot \langle \vec{P} \rangle }{b(\vec{a})} ] ,
\label{magicmeasurer}
\end{align}
where the maximization is performed over all $ \vec{a} $.  Since $ \langle P_0 \rangle = 1 $ for any state, we may take $ a_0 = 0 $ leaving the remaining $ D^2 - 1$ variables to be optimized.  For $ \vec{a} = \vec{0} $, we take the argument of the maximization to be 1, which guarantees that $ {\cal M} (\rho) \ge 1 $.  The quantity to be maximized is (\ref{polytopeboundary}), such if there is any hyperplane which shows a violation, we will have $ {\cal M}(\rho) >1 $.
This is a necessary and sufficient criterion for magic when $ {\cal M} (\rho) >1 $.  The quantifier defined above possesses key properties that make it a valid monotone \cite{streltsov2017colloquium}: 1) $ {\cal M} (\rho) \ge 1 $;
2) Invariance under Clifford unitaries $ {\cal M} (\rho) = {\cal M} (U_C \rho U_C^\dagger ) $; 3) Faithfulness $ {\cal M} (\rho) = 1 $ iff $ \rho = \rho_S$, otherwise $ {\cal M} (\rho) > 1 $; 4) Monotonicity  $ {\cal M} ( {\cal E} (\rho)) \le {\cal M} ( \rho) $, where $ {\cal E} $ is a stabilizer channel; 5) Convexity $ {\cal M} ( \sum_k p_k \rho_k ) \le \sum_k p_k {\cal M} ( \rho_k ) $ (see Appendix).  

The definition (\ref{magicmeasurer}) shows how the magic states are distributed in $ \cal P$-space.  Consider the set of states with equal magic $ {\cal M} (\rho) = r $.  This defines a set of hyperplanes $ \vec{a} \cdot \langle \vec{P} \rangle = r b(\vec{a})  $, which corresponds to an enlarged polytope that has the same shape as the stabilizer polytope (Figure \ref{fig1}(a)). The form of (\ref{magicmeasurer}) is consistent with a Minkowski functional \cite{Narici2010} which is a way of measuring the distance of a point from the stabilizer polytope by seeing how much the polytope has to be scaled up in order for it to just include the point. As a result, this measure inherits the same symmetries as the underlying stabilizer polytope.

An interesting point here is that this structure precludes alternative definitions of monotones based on, for instance, Euclidean distance of states in $ \cal P $-space from the polytope. Such a measure would result in a polytope with rounded edges and corners when finding points with constant magic, which is 
inconsistent with points of constant RoM even for $ N = 1$. It would also not be comparable without rescaling between points whose nearest hyperplanes belong to different families for multi-qubit systems, due to Euclidean distance being spherically symmetric.

\paragraph{Numerical demonstration}

We now show some explicit numerical examples using our methods to show its utility in a mixed state context. In addition to explicitly calculating $ {\cal M}(\rho) $, we also use the necessary and sufficient conditions (\ref{allpolytopeboundaries}) to  construct a witness
\begin{align}
{\cal W }(\rho)  = \max_{ \{ |a_k| \le 1 , a_0 = 0  \}}  [ \vec{a} \cdot \langle \vec{P} \rangle  -  b(\vec{a})  ] ,
\label{magicmeasuresimple}
\end{align}
which returns a positive value $ {\cal W }(\rho) > 0 $ for states with magic. Here, we normalized the vector $ \vec{a} $ such that all coefficients lie in the range $ |a_k | \le 1 $ (see Appendix). While this means that strictly $ a_k $ takes only rational values, numerically there is little benefit of this constraint and we treat $ a_k $ as a real parameter.  
We find that for the small scale systems as plotted here, the maximization for both $ {\cal M}(\rho) $ and $ {\cal W}(\rho) $ can be calculated within one minute with modest computational resources. 
We find evaluating our quantities is much faster than evaluating RoM, which involves a nonlinear optimization over $ D_S \sim 2^{N^2/2} $ real variables.  While RoM is in principle a faithful measure of magic, the large computational overhead effectively can give false positives as a magic detector, due to the imperfect optimization giving a decomposition with negative coefficients.

\begin{figure}[t]
\includegraphics[width=\linewidth]{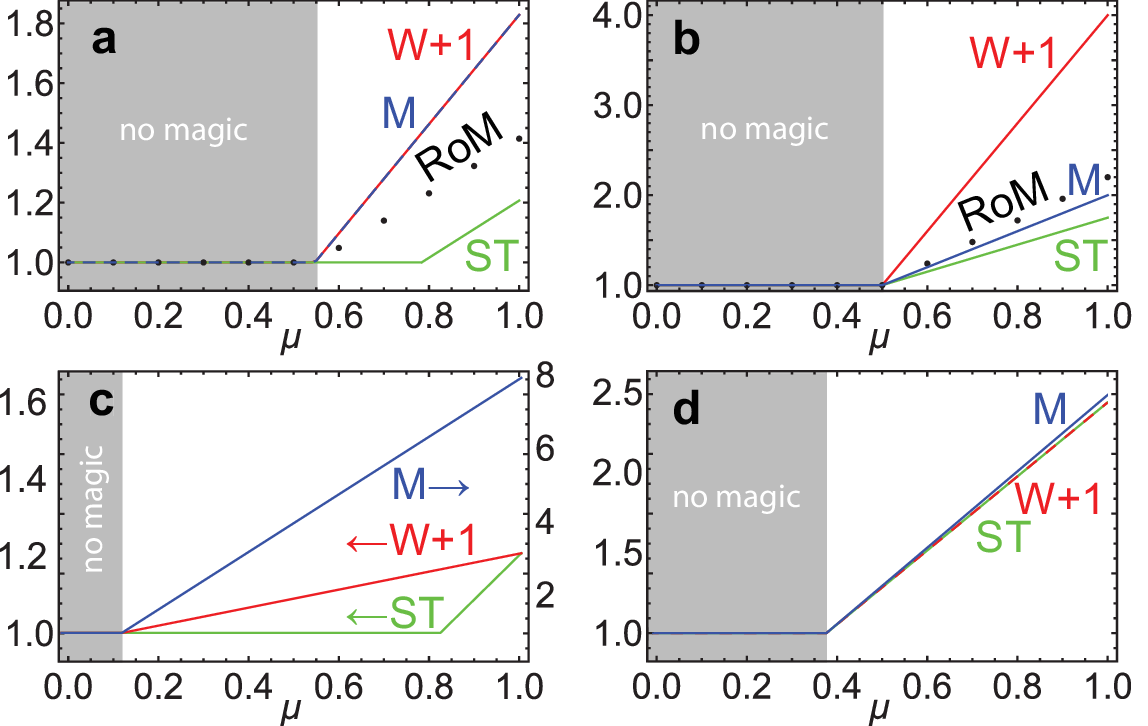}
\caption{Comparison of various magic quantifiers for various Werner states defined as $ \rho = (1- \mu ) I/D + \mu |\psi\rangle \langle \psi | $. Calculated are states (a) $ |\psi \rangle = (|00 \rangle + e^{i \pi/4} |11 \rangle )/\sqrt{2} $; 
(b) $ |\psi \rangle = (|00 \rangle + |01 \rangle +|10 \rangle + i |11 \rangle )/2 $; (c) $ |\psi \rangle = (|00000 \rangle + e^{i \pi/4} |11111 \rangle )/\sqrt{2} $; (d) $ |\psi \rangle = (|0\rangle + e^{i \pi/4} |1 \rangle)^{\otimes 5} /\sqrt{2^5} $. Lines shown are Robustness of magic (RoM), stabilizer norm (ST), our magic monotone (\ref{magicmeasurer}) (M), and magic witness (\ref{magicmeasuresimple}) (W+1). In (c) the arrows indicate the axes that are used for each quantity.  For all plotted quantities a value greater than 1 indicates the presence of magic. The magic witness $ \cal W (\rho) $ we have added 1 to (\ref{magicmeasuresimple}).  For the stabilizer norm we plot $ \text{ST} = \max ( 1, || \rho ||_{\text{st}} ) $.
\label{fig2}  }
\end{figure}

Figure \ref{fig2} shows a comparison of various quantifiers for various Werner states.  We see that both our magic witness and monotone successfully detects magic in the same region as RoM for $ N = 2$ (Fig. \ref{fig2}(a)(b)) and gives consistent results for $ N = 5$ (Fig. \ref{fig2}(c)(d)). Both of these quantities often shows an improvement in the detection range over the stabilizer norm, which is only a sufficient condition for magic.  Examining the expression for the stabilizer norm (\ref{stnorm}), we can see that this is a particular case of our criterion where $ \vec{a} = \text{sgn}(\langle P_n \rangle ) $ and $ b(\vec{a}) = 2^N $.  This would corresponding one particular choice of hyperplane, which may not correspond to the polytope boundary giving the tightest bound.  By running over all polytope boundaries, our witness is able to detect magic states that are missed by the stabilizer norm.

\paragraph{Conclusions}

We have introduced an approach to detect and quantify the amount of magic in an arbitrary quantum state by finding the hyperplane equations defining the stabilizer polytope.  By testing all possible hyperplanes, one can obtain a necessary and sufficient criterion for detecting magic.  This can be adapted into a magic monotone which quantifies the amount of magic according to the scale factor required to enlarge the polytope such that it falls on its boundary.  We find that the approach works well numerically, where magic can be detected in mixed states much more efficiently than other faithful mixed state monotones such as RoM.  There are numerous ways that this approach can be developed further, and thereby improving methods for magic detection.  A better understanding of the polytope boundaries for a given $ N $ would allow one to further constrain the maximization in (\ref{magicmeasure}), to reduce the search space of the hyperplanes.  Improvements in obtaining the bound 
$ b(\vec{a} ) $, which in a brute force approach involves a discrete maximization over $ D_S $, would lead to further improvements in efficiency, since the remaining optimization in (\ref{magicmeasure}) involves a smaller $D^2-1$ variables.  By further developing these techniques it is likely that the magic in larger systems can be quantified more efficiently.


This work is supported by the National Natural Science Foundation of China (62071301); NYU-ECNU Institute of Physics at NYU Shanghai; the Joint Physics Research Institute Challenge Grant; the Science and Technology Commission of Shanghai Municipality (19XD1423000,22ZR1444600); the NYU Shanghai Boost Fund; the China Foreign Experts Program (G2021013002L); the NYU Shanghai Major-Grants Seed Fund; Tamkeen under the NYU Abu Dhabi Research Institute grant CG008; and the SMEC Scientific Research Innovation Project (2023ZKZD55).

 

\providecommand{\noopsort}[1]{}\providecommand{\singleletter}[1]{#1}%

\appendix

\section{Stabilizers in $ \cal P $-space}

Here we show that the stabilizers in $ \cal P$-space $ \vec{S}_i = \langle S_i | \vec{P} | S_i \rangle $ have $ 2^N $ non-zero elements, corresponding to $ 2^N $ mutually commuting Pauli strings. 

First note that any stabilizer state can be written 
\begin{align}
    |S_i \rangle = U_C | 0 \rangle^{\otimes N }
\end{align}
since any pure stabilizer state can be obtained by a Clifford unitary rotation from the state $ | 0 \rangle^{\otimes N } $.  The stabilizer vector can be written 
\begin{align}
\vec{S}_i & = \text{Tr} ( |S_i \rangle \langle S_i | \vec{P} ) \\
& =  \text{Tr} [ U_C (| 0 \rangle \langle 0 | )^{\otimes N } U_C^\dagger  \vec{P} ]  \\
 & =   \frac{1}{2^N}  \text{Tr} [  U_C [ \bigotimes_{n=1}^N ( I_n + Z_n) ] U_C^\dagger \vec{P} 
 ]  \\
 & =  \frac{1}{2^N}  \text{Tr} [  U_C [ I^{\otimes N} + \sum_{n=1}^N Z_n + \dots + \bigotimes_{n=1}^{N} Z_n ) ] U_C^\dagger \vec{P} ] \label{zexpansion}
\end{align}
where in the third line we used the fact that $ | 0 \rangle \langle 0 | = (I+Z)/2 $.  The last line is a sum over $ 2^N $ Pauli strings.  The Clifford unitary  will transform these to another sum of $ 2^N $ Pauli strings (up to a $ \pm 1 $ factor).  Using the orthogonality of Pauli strings $ \text{Tr} (P_n P_m ) = 2^N \delta_{nm} $, we obtain the stated result.

The non-zero elements correspond to mutually commuting Pauli strings since the $ 2^N $ terms in (\ref{zexpansion}) are Clifford transformations of tensor products of $ Z_n $ operators, which are mutually commuting.

\section{Bounds of the hyperplane}

Here we prove that (\ref{polytopeboundary}) gives the general form of the boundary of the polytope.   Substituting (\ref{polytopevec}) into the left hand side of (\ref{polytopeboundary}), we have
\begin{align} 
\min_{i\in[1,D_S] } \vec{a} \cdot \vec{S}_i  \le \sum_{i=1}^{D_S} p_i \vec{a} \cdot  \vec{S}_i \le \max_{i\in[1,D_S] } \vec{a} \cdot \vec{S}_i 
\label{lowerupperbound}
\end{align}
which follows from the fact that $ 0\le p_i \le 1 $ and $\sum_i p_i =1  $ is a probability.   The inequality for the upper bound is (\ref{polytopeboundary}).  

The lower bound can be rewritten in the same form as the upper bound by redefining $ \vec{a} \rightarrow - \vec{a} $
\begin{align} 
\min_{i\in[1,D_S] } -\vec{a} \cdot \vec{S}_i &  \le - \sum_{i=1}^{D_S} p_i \vec{a} \cdot  \vec{S}_i  \\
-\max_{i\in[1,D_S] } \vec{a} \cdot \vec{S}_i &  \le - \sum_{i=1}^{D_S} p_i \vec{a} \cdot  \vec{S}_i \\
\max_{i\in[1,D_S] } \vec{a} \cdot \vec{S}_i &  \ge  \sum_{i=1}^{D_S} p_i \vec{a} \cdot  \vec{S}_i 
\end{align}
which is the same as (\ref{polytopeboundary}). We may therefore use (\ref{lowerupperbound}) for a single choice of $ \vec{a}$, or use (\ref{polytopeboundary}) with with both $ \vec{a}, - \vec{a} $.

\section{Finding a polytope boundary}

We first show that (\ref{acond}) and (\ref{polytopecond}) are the two equations that need to be satisfied for a polytope boundary. A hyperplane running through a face $ F $ with coordinates $\{ \vec{S}_1^F, \dots , \vec{S}_{D_F}^F \} $ must satisfy
\begin{align}
\vec{a} \cdot \vec{S}_j^F =  b
\label{hyperplanebs}
\end{align}
for all $ j \in [1,D_F] $. Subtracting these equations yields (\ref{acond}). 

In order for this to be a polytope boundary, $ b $ must additionally satisfy (\ref{bcond}).  We thus require
\begin{align}
\vec{a} \cdot \vec{S}_1^F = \max_{i\in[1,D_S] } \vec{a} \cdot \vec{S}_i 
\end{align}
where we have taken $ j = 1 $ in (\ref{hyperplanebs}) since all $ j $ give the same $ b $. We may equally write this as a inequality running over $ i \in [1,D_S] $
\begin{align}
    \vec{a} \cdot \vec{S}_1^F \ge \vec{a} \cdot \vec{S}_i ,
\end{align}
which is equivalent to (\ref{polytopecond}).

Eqs. (\ref{acond}) and (\ref{polytopecond}) together give a method to determine whether a given set $F $ is a polytope boundary.  The procedure can be summarized as
\begin{enumerate}
\item Pick a set of stabilizers $F $ with coordinates $\{ \vec{S}_1^F, \dots , \vec{S}_{D_F}^F \} $ 
\item Solve (\ref{acond}) and determine $ \vec{a} $.  For an underconstrained problem this may leave free parameters.  Write $ \vec{a} $ in terms of these free parameters. If there is no solution of (\ref{acond}), then return false.  
\item Solve (\ref{polytopecond}) and determine $ \vec{a} $ using the parameterization from step 2.  
\item If a solution of (\ref{polytopecond}) can be found, return $ \vec{a} $ and $ b = \vec{a} \cdot \vec{S}_1^F $ which gives the hyperplane of the polytope boundary.  If no solution is found, then return false. 
\end{enumerate}

As an example, we give details of the computation of finding a polytope boundaries in Fig. \ref{fig1}(a) for $ N = 1$. Choosing $ F = \{ |+ \rangle, | i \rangle \}  $ gives the stabilizer vectors $ \vec{S}^F = \{ (1,1,0,0), (1,0,1,0) \} $.  From (\ref{acond}), we require $ \vec{a} = (0,1,1, a_3 ) $, which sets $ b = 1 $, up to a common multiplicative factor.  In order to satisfy (\ref{polytopecond}) we require $ -1 \le  a_3 \le 1 $.  This is an example of an underconstrained problem, where the hyperplane can take a range of values.  Choosing $ a_3  = 0 $ gives the boundary 
\begin{align}
\langle X \rangle + \langle Y \rangle \le 1
\label{qubitxy}
\end{align}
as shown in Fig.  \ref{fig1}(a).  Other choices of $ a_3 $ equally form a valid polytope boundary.  In this case, adding another stabilizer  $ |0 \rangle $ to $ F $ fully constrains the parameters of the hyperplane to give 
\begin{align}
\langle X \rangle + \langle Y \rangle+ \langle Z \rangle  \le 1 ,
\end{align}
which gives one of the faces of the stabilizer polytope.  Other choices of stabilizers give the remaining faces of the octahedron (\ref{octahedron}).

\section{Symmetries of the polytope}

\paragraph{Clifford symmetry}

Under a Clifford transformation, the polytope face (\ref{polytopeboundaryrho}) becomes
\begin{align}
\rho_F \rightarrow \rho_{F'} = U_C \rho_F U_C^\dagger & = \sum_{j=1}^{D_F} p_j U_C |S_j^F \rangle \langle S_j^F | U_C^\dagger \\
& = \sum_{j=1}^{D_F} p_j |S_{C(j)} \rangle \langle S_{C(j)}| 
\label{facetransformclifford}
\end{align}
where $ C(j) \in [1, D_S ]$ is a permutation of the stabilizer labels according to the Clifford transformation $ U_C $.  We then may define the transformed face $ F' $ as the subset of stabilizers consisting of $ |S^{F'}_j \rangle = |S_{C(j)} \rangle$.  

Suppose now we have found a valid polytope boundary for the face $ F $, given by (\ref{polytopeboundary}). Then the hyperplane for the face $ F' $ satisfies
\begin{align}
\vec{a}' \cdot  \langle \vec{P} \rangle = b(\vec{a}') 
\end{align}
where $ \vec{a}' $ are the new coefficients.  We may deduce what these coefficients are by a Clifford transformation of (\ref{polytopeboundary}), which gives  
\begin{align}
\vec{a} \cdot  \langle \vec{P}' \rangle = b(\vec{a}),
\label{transformedhyper}
\end{align}
%
%
where we defined $ \vec{P}' = U_C^\dagger  \vec{P} U_C$, which is also a vector of Pauli strings due to the property of Clifford transformations.  The ordering of the Pauli strings will not be in the canonical order of $ \vec{P} $ and may have additional $ -1 $ factor on the elements, hence we have
\begin{align}
    P_k' = \pm  P_{c(k)} ,
    \label{paulitransform}
\end{align}
where $ c(k) \in [1,D^2] $ is a permutation function. The coefficients $ \vec{a} ' $ must then be
\begin{align}
    a_k' = \pm a_{c(k)} 
\end{align}
where the $ \pm $ factor is the same as in (\ref{paulitransform}). 

The constant factor in the hyperplane (\ref{polytopeboundary}) is unchanged. 
 To see this, first note that the $ \cal P $-space coordinates of the stabilizers transform as
 \begin{align}
\vec{S}_i \rightarrow \vec{S}_{C(i)} & = \langle S_i | \vec{P}' | S_i \rangle = \vec{S}_i' ,
\end{align}
which have the same permutation of its vector elements up to a $ \pm 1$ factor as (\ref{paulitransform}).  Now we may write
\begin{align}
b( \vec{a}' ) & = \max_{i\in[1,D_S] } \vec{a}' \cdot \vec{S}_i \\
& = \max_{i\in[1,D_S] } \vec{a}' \cdot \vec{S}_{C(i)} \\
& = \max_{i\in[1,D_S] } \vec{a}' \cdot \vec{S}_{i}'
\end{align}
since the maximum runs over all stabilizers.  Since both $ \vec{a} '$ and $ \vec{S}_i'$ have the same permutation and $ \pm 1 $ factors, the dot product evaluates to be the same and we conclude that $ b(\vec{a}') = b(\vec{a}) $.

\paragraph{Reflection symmetry}

The transformation 
\begin{align}
    P_n^{(l)} \rightarrow - P_n^{(l)} ,  l \in [1,3] 
    \label{spinfliprule}
\end{align} 
is another type of stabilizer preserving transformation.  To see this, note  that a stabilizer state $ |S_i \rangle $ is the simultaneous eigenstate of $ 2^N $ commuting Pauli strings $ P_n $. Given $ 2^N $ commuting Pauli strings, the number of eigenstates is $ 2^N $.  For example, for the set of commuting Pauli strings $ \{ I, Z_1, Z_2, Z_1 Z_2 \} $, the eigenstates are $ \{ |00\rangle, |01\rangle, |10\rangle, |11\rangle \} $. Adjusting the signs on the Pauli strings $ P_n $ and demanding that the stabilizer is the $ +1 $ eigenvalue narrows it down to a single stabilizer state. Since the transformation (\ref{spinfliprule}) only changes the sign of $ 4^{N-1} $ of the $ P_n $, this then specifies another stabilizer state.  Note that this can only be done according to the rule (\ref{spinfliprule}), and the signs of $ P_n $ cannot be changed arbitrarily.  For example, for the set of Pauli strings $ \{ I, -Z_1, Z_2, Z_1 Z_2 \} $, there is no eigenstate with +1 eigenvalue for all Pauli strings.  

We define this symmetry by first writing
\begin{align}
|S_i \rangle \langle S_i | & = \frac{1}{2^N} \vec{S_i} \cdot \vec{P} 
\end{align}
where we used (\ref{zexpansion}) and the fact that any density matrix can be expanded in terms of the Pauli strings $ \rho = \langle \vec{P} \rangle \cdot \vec{P} /2^N  $.  Then under (\ref{spinfliprule}) the stabilizer transforms as
\begin{align}
|S_i \rangle \langle S_i | \rightarrow  R[|S_i \rangle \langle S_i |] & = \frac{1}{2^N} \vec{S}_i \cdot \vec{P}'' \\
& = \frac{1}{2^N} \vec{S}_{R(i)} \cdot \vec{P} \label{signchange} \\
& =|S_{R(i)} \rangle \langle S_{R(i)} | 
\end{align}
where $ \vec{P} '' $ has sign changes according to (\ref{spinfliprule}), and  in (\ref{signchange}) we transferred the additional minus signs from $   \vec{P}''$ to $ \vec{S}_i $ in the dot product. $ R(j) \in [1,D_S ]$ is a permutation of the stabilizer labels.  Then similarly to (\ref{facetransformclifford}), we have 
\begin{align}
\rho_F \rightarrow \rho_{F''}  = R[\rho_F] =  \sum_{j=1}^{D_F} p_j |S_{R(j)} \rangle \langle S_{R(j)}| ,
\label{facetransformreflection}
\end{align}
which is another polytope face. 

Similarly to (\ref{transformedhyper}), if we have a valid polytope boundary for $ F $, then 
\begin{align}
\vec{a} \cdot  \langle \vec{P}'' \rangle= b(\vec{a}) 
\end{align}
where $ \vec{P} '' $ is the transformed Pauli string vector with additional minus signs is also a polytope boundary.

\section{General form of the hyperplane parameters}

We show here that it is sufficient to take $ \vec{a} $ with integer elements.  Following the discussion in the main text, a polytope boundary must satisfy the relations (\ref{acond}) and (\ref{polytopecond}).  Depending upon the number of stabilizers in $ F $, this may result in an underconstrained set of equations.  In this case, it is of course possible to take $ \vec{a} $ with non-integer coefficients due to the presence of unconstrained variables, as explicitly shown for the qubit case in the discussion surrounding (\ref{qubitxy}).  Such underconstrained hyperplanes do not give the tightest bounds to the polytope. For example, in the qubit case such solutions would give boundaries that are between the hyperplanes indicated in Fig. \ref{fig1}(a).  In order to most effectively bound the polytope, we will be interested in the nature of the coefficients $ \vec{a} $ for the fully constrained case, where $ F $ contains enough stabilizers such that it leaves no free parameters.  

We now introduce another way to write (\ref{acond}) to find $ \vec{a} $.  First let us define a linearly independent subset of the stabilizer vectors in $ F $
\begin{align}
F_{\text{ind}} = \{ \vec{S}^{F_{\text{ind}}}_1, \vec{S}^{F_{\text{ind}}}_2, \dots , 
\vec{S}^{F_{\text{ind}}}_{D^2-1}  \}
\end{align}
Here there are $ D^2-1 $ vectors, which is the maximum dimension that can be spanned by the stabilizer vectors, since the first element of any stabilizer state is $ [\vec{S}_{i} ]_0 = \langle I \rangle = 1 $.  We define $ \vec{S}^{F_{\text{ind}}}_i $ omitting this first element, and thus it is a $ D^2-1 $ dimensional vector.  These are chosen from the set $ F $, removing any linearly dependent vectors.  We now form the matrix
\begin{align}
    A= (\vec{S}^{F_{\text{ind}}}_1, \vec{S}^{F_{\text{ind}}}_2, \dots , 
\vec{S}^{F_{\text{ind}}}_{D^2-1}  )^T
\end{align}
where the rows of the matrix are the stabilizer vectors in $ F_{\text{ind}} $.  
Then the hyperplane equation satisfies
\begin{align}
A \vec{a} = c \vec{1}
\label{matrixmethod}
\end{align}
where $\vec{1} = (1,1,\dots,1)^T $ and $ c $ is a constant.  Then to find $ \vec{a} $ we must invert $ A$, which gives the solution
\begin{align}
\vec{a} = c A^{-1} \vec{1} . 
\label{invmatrixmethod}
\end{align}
Since $ A $ is a full rank matrix, the inverse exists.  

In this way of solving for $ \vec{a} $, one can see why $ \vec{a} $ only takes integer elements.  Since $ A $ is a matrix that consists of stabilizer vectors, according to (\ref{zexpansion}) it only contains elements $ \{ 0, \pm 1 \} $.  This is a simple example of a rational matrix.  Since the inverse of a rational matrix is also a rational matrix, $ A^{-1} $ involves only rational elements.  Therefore from (\ref{invmatrixmethod}), $\vec{a} $ will take rational elements multiplied by $ c $. Since the equation of a hyperplane (\ref{hyperplane}) can be multiplied by an overall constant without change, we may choose $c$ as the lowest  common denominator of the rational $ a_k $ without loss of generality, which converts $ \vec{a} $ to an integer vector. The hyperplane intercept $ b $ then can be determined from (\ref{bcond}), which obviously gives an integer value since $ \vec{S}_i $ is an integer vector.

\section{Necessary and sufficient conditions}

Consider an arbitrary state given by $ \rho $.  In $ \cal P $-space this can be written by the vector $ \langle \vec{P} \rangle = \text{Tr} (\rho \vec{P}) $.  Now consider the set of all possible polytope boundaries (\ref{allpolytopeboundaries}).  We claim that a violation of (\ref{allpolytopeboundaries}) is a necessary and sufficient condition for a magic state. 

For the sufficient condition, note that (\ref{polytopeboundary}) is a true statement for any stabilizer mixture (\ref{mixedstabilizer}).  To see this, subtitute (\ref{mixedstabilizer}) and (\ref{bcond}) into (\ref{polytopeboundary}), where we obtain
\begin{align}
\sum_{i=1}^{D_S} p_i \vec{a} \cdot S_i \le \max_{i\in[1,D_S] } \vec{a} \cdot \vec{S}_i  .
\end{align}
This is true from the fact that $ 0 \le p_i \le 1 $ and $ \sum_i p_i = 1 $ is a probability.  Since (\ref{mixedstabilizer}) is the most general form of a mixed stabilizer state, any violation of (\ref{polytopeboundary}) must arise from the fact that the state is not a stabilizer state.  Since a violation of any of the inequalities (\ref{polytopeboundary}) would constitute a violation of (\ref{allpolytopeboundaries}), this is a sufficient condition for detecting magic. 

Now we prove that a violation of (\ref{allpolytopeboundaries}) is also a necessary condition for a magic state.  If (\ref{allpolytopeboundaries}) is to be a necessary condition, then we require that there are no states that satisfy (\ref{allpolytopeboundaries}), yet possesses magic. We prove this by contradiction.  Suppose there is a state $ \rho_{\text{test}} $ with $ \langle \vec{P} \rangle_{\text{test}} = \text{Tr} (\rho_{\text{test}} \vec{P}) $ that satisfies all the bounds in (\ref{allpolytopeboundaries}), but in fact possesses magic.  First note that the set of all half-planes in (\ref{allpolytopeboundaries}) defines a convex polytope \cite{grunbaum1967convex}.  Since
$ \rho_{\text{test}} $ satisfies all the bounds (\ref{allpolytopeboundaries}), it must be either on the boundary or within the polytope.  However, any point within the convex polytope can be written
\begin{align}
\langle \vec{P} \rangle = \sum_{i=1}^{D_S} p_i \vec{S}_i ,  
\end{align}
with $ 0\le p_i \le 1 $ and $ \sum_i p_i = 1 $.  This is a zero magic state and therefore we have arrived at a contradiction. This shows that any magic state must violate (\ref{allpolytopeboundaries}), meaning that it is a necessary condition.

\section{Properties of the magic monotone $ {\cal M}(\rho) $}

In this section we prove the following properties of the monotone $ {\cal M}(\rho) $: 
1) $ {\cal M} (\rho) \ge 1 $;
2) Invariance under Clifford unitaries $ {\cal M} (\rho) = {\cal M} (U_C \rho U_C^\dagger ) $; 3) Faithfulness $ {\cal M} (\rho) = 1 $ iff $ \rho = \rho_S$, otherwise $ {\cal M} (\rho) > 1 $; 4) Monotonicity  $ {\cal M} ( {\cal E} (\rho)) \le {\cal M} ( \rho) $, where $ {\cal E} $ is a stabilizer channel; 5) Convexity $ {\cal M} ( \sum_k p_k \rho_k ) \le \sum_k p_k {\cal M} ( \rho_k ) $.

\paragraph{Greater or equal to 1}

The quantity $ {\cal M} (\rho) $ is greater or equal to 1 since $ \vec{a} = \vec{0} $ is always a candidate hyperplane in the maximization over $ \vec{a} $.  This hyperplane always gives 
\begin{align}
\vec{a} \cdot \langle \vec{P} \rangle =b (\vec{a}) =  0 
\label{nonneg}
\end{align}
hence the numerator and denominator in (\ref{magicmeasurer}) is equal to zero.  The argument of the maximization (\ref{magicmeasurer}) is indeterminate as written, which we take to be equal to 1 for this special point since it can be considered a trivial hyperplane where $ \vec{a} \cdot \langle \vec{P} \rangle = b (\vec{a}) $ is always satisfied.  

Since the maximization may potentially attain a larger value for another hyperplane, we have $ {\cal M} (\rho) \ge 1 $.

\paragraph{Invariance under Clifford transformations}

We first write the monotone (\ref{magicmeasurer}) as
\begin{align}
{\cal M} ( \rho ) = \frac{ \vec{a}_{\text{opt}} \cdot \langle \vec{P} \rangle_\rho   }{ b(\vec{a}_{\text{opt}} ) } ,
\end{align}
where $ \vec{a}_{\text{opt}}  $  is the result of the maximization in (\ref{magicmeasure}). 

Under a Clifford transformation, the coordinates of any state in $ \cal P $-space transform as
\begin{align}
\langle \vec{P} \rangle_\rho  \rightarrow \langle \vec{P} \rangle_{U_C \rho U_C^\dagger } & = \text{Tr} ( \vec{P} U_C \rho U_C^\dagger ) \\
& = \text{Tr} ( U_C^\dagger \vec{P} U_C \rho  ) = \langle \vec{P}' \rangle_\rho . 
\label{cliffordtranspvec}
\end{align}
Since a Clifford transformation turns a Pauli string into another Pauli string (up to a sign), $ \langle \vec{P} \rangle $ and $\langle \vec{P} ' \rangle  $ are related by a permutation of its elements, up to an multiplicative $ \pm 1$ factor.  

Under a Clifford transformation, the stabilizer states map to another stabilizer state $ |S_i \rangle \rightarrow U_C |S_i \rangle \propto |S_{C(i)} \rangle $, where $C(i) $ is a permutation operation.  The vectors in $ \cal P $-space map as
\begin{align}
\vec{S}_i \rightarrow \vec{S}_{C(i)} & = \langle S_i | U_C^\dagger \vec{P} U_C | S_i \rangle   \\
& = \langle S_i | \vec{P}' | S_i \rangle = \vec{S}_i'
\end{align}
hence is again related by a permutation of the elements up to a $ \pm 1 $ sign.  

Now note that 
\begin{align}
b( \vec{a} ) & = \max_{i\in[1,D_S] } \vec{a} \cdot \vec{S}_i \\
& = \max_{i\in[1,D_S] } \vec{a} \cdot \vec{S}_{C(i)} \\
& = \max_{i\in[1,D_S] } \vec{a} \cdot \vec{S}_{i}'
\label{boptvarious}
\end{align}
since the maximum runs over all stabilizers.  

Then we may write 
\begin{align}
{\cal M} ( U_C \rho U_C^\dagger ) =\frac{\vec{a}_{\text{opt}}' \cdot \langle \vec{P}' \rangle    }{\max_{i\in[1,D_S] } \vec{a}_{\text{opt}}'  \cdot \vec{S}_{i}' } ,
\end{align}
where $ \vec{a}_{\text{opt}} ' $  is the result of the maximization for the transformed state. Since both $ \langle \vec{P} ' \rangle $ and $ \vec{S}_i'$ are related to the original $ \langle \vec{P} \rangle $ and $ \vec{S}_i$ by the same permutation and sign changes, the optimal  $ \vec{a}_{\text{opt}} ' $ is related to the original  $ \vec{a}_{\text{opt}}  $ by the same permutation and sign changes.  The dot product is invariant under permutation and sign changes of the components $ \vec{a} \cdot \langle \vec{P} \rangle = \vec{a}' \cdot \langle \vec{P} '\rangle $.  The vector norm is also invariant under permutation and sign changes of the components $ || \vec{a}_{\text{opt}} || = || \vec{a}_{\text{opt}}' || $. We therefore conclude that 
\begin{align}
{\cal M} ( \rho ) = {\cal M} ( U_C \rho U_C^\dagger ) .
\end{align}


\paragraph{Faithfulness}

We have already shown that for any stabilizer mixture, 
\begin{align}
\vec{a}_{\text{opt}} \cdot \langle \vec{P} \rangle_{\rho_S}   \le b(\vec{a}_{\text{opt}} ) 
\label{stabcond}
\end{align}
from (\ref{polytopeboundary}). Hence for any choice of $ \vec{a} $, the quantity inside the maximization (\ref{magicmeasurer}) is less than or equal to 1 for a stabilizer mixture.  From the same arguments showing that ${\cal M} (\rho) \ge 1 $, it follows that ${\cal M} (\rho_S ) \ge 1 $. The only consistent value for a stabilizer mixture is then ${\cal M} (\rho_S ) =1 $.  

Furthermore, we have shown that violation of (\ref{allpolytopeboundaries}) is a necessary and sufficient condition for the detection of magic.  Then all magic states violate (\ref{stabcond}) and give a value ${\cal M} (\rho ) > 1 $.

\paragraph{Monotonicity}

First let us define the concept of the $ r$-polytope.  Consider the set of all points in $ \cal P $-space that are a convex sum of stabilizer vectors rescaled by a factor $ r \vec{S}_i $:
\begin{align}
\vec{R} = r \sum_{i=1}^{D_S} p_i \vec{S}_i ,
\label{rpolytope}
\end{align}
where $ r \ge 1 $, $  0 \le p_i \le 1 $, and $ \sum_i p_i = 1 $. The set of all possible $ \vec{R} $ forms a convex polytope that is scaled up by a factor $ r $, and the vectors in the conventional stabilizer polytope (\ref{polytopevec}) corresponds to $ r = 1$.  It follows that for two such polytopes with $ r, r' $, if $ r' < r $, then all points in the $ r' $-polytope are contained within the $ r $-polytope.

Since $ \vec{S}_i $ forms a complete non-orthogonal basis except for the first element which is always $ a_0 = 1 $, by choosing $ r $ sufficiently large, an arbitrary state $ \langle \vec{P} \rangle_\rho  $ can be written in the form (\ref{rpolytope}).  Let us then choose an $ r $ such that $ \langle \vec{P} \rangle_\rho  $, lies on one of the faces of the $ r$-polytope
\begin{align}
\langle \vec{P} \rangle_\rho = r \sum_{i=1}^{D_F} p_i \vec{S}_i^F .
\label{rhostate}
\end{align}
Now let the hyperplane defining this face have the parameters $ \vec{a}^F $ from which we may deduce that the 
\begin{align}
\vec{a}^F \cdot \langle \vec{P} \rangle_\rho & = r \vec{a}^F \cdot  \sum_{i=1}^{D_F} p_i \vec{S}_i^F \label{hyperface}  \\
& = r b(\vec{a}^F)  \sum_{i=1}^{D_F} p_i = r b(\vec{a}^F)
\label{monohyper}
\end{align}
where we used (\ref{hyperplane}) and (\ref{bcond}).  We may then evaluate (\ref{magicmeasurer}) as
\begin{align}
{\cal M} (\rho) = r .
\label{Risr}
\end{align}

Now consider a trace-preserving stabilizer preserving channel \cite{howard2017application,seddon2021quantifying} which obeys 
\begin{align}
    {\cal E}(|S_i \rangle \langle S_i | ) = \sum_j P^{({\cal E})}_{ij }|S_j \rangle \langle S_j| 
    \label{stabchannel}
\end{align}
where $ 0 \le P^{({\cal E})}_{ij } \le 1  $ is a probability distribution satisfying $ \sum_j P^{({\cal E})}_{ij } = 1$.  The state (\ref{rhostate}) then transforms as
\begin{align}
\langle \vec{P} \rangle_{{\cal E}(\rho)} = r \sum_{i=1}^{D_F} \sum_{j=1}^{D_S} p_i P^{({\cal E})}_{ij } \vec{S}_j  .
\label{rhostateepsgen}
\end{align}
We may write this equivalently as 
\begin{align}
\langle \vec{P} \rangle_{{\cal E}(\rho)} = r \sum_{j=1}^{D_S} p_j'  \vec{S}_j  .
\label{stabchanp}
\end{align}
where 
\begin{align}
   p_j'=  \sum_{i=1}^{D_F} p_i P^{({\cal E})}_{ij },
\end{align}
is a probability distribution with $ \sum_j p_j' = 1$.  Eq. (\ref{stabchanp}) describes the coordinates of a point within the $ r$-polytope, since it is a convex sum of rescaled stabilizer vectors $ r \vec{S}_i $.  

For the case that (\ref{stabchanp}) lies on the surface of the $ r$-polytope, similar arguments to (\ref{rhostate})-(\ref{Risr}) lead to the conclusion that 
\begin{align}
    {\cal M} ({\cal E}(\rho)) = r.
    \label{Rer}
\end{align}  
For the general case $ \langle \vec{P} \rangle_{{\cal E}(\rho)}  $ lies within the $ r $-polytope, we may construct a smaller $ r'$-polytope such that the state lies on the face of it as
\begin{align}
\langle \vec{P} \rangle_{{\cal E}(\rho)} = r' \sum_{i=1}^{D_{F'}} p_i \vec{S}_i^{F'} .
\label{rhostate}
\end{align}
Then according to similar arguments to (\ref{monohyper})-(\ref{Risr}), we conclude that 
\begin{align}
    {\cal M} ({\cal E}(\rho)) = r'.
        \label{Rerp}
\end{align}  
Combining (\ref{Rer}) and (\ref{Rerp}) and the fact that $ r' < r $, we have 
\begin{align}
{\cal M} ({\cal E} (\rho) ) \le  r  = {\cal M} (\rho) 
\end{align}
which shows monotonicity.



\paragraph{Convexity}

Let us consider a linear space $X$ with an absorbing subset $C$. The Minkowski functional of $C$, \\
$P: X \rightarrow [0, \infty]$ is defined as
\begin{equation}
    P(x) = \inf \left \{ \lambda > 0 : x \in \lambda C \right\}
\end{equation}
and acts to find by how much the absorbing subset $C$ has to be scaled up in order to just contain the point $x$. 

The stabilizer polytope contains the origin and spans ${\cal P}$-space making it an absorbing subset of ${\cal P}$-space, and hence allowing us to define the Minkowski functional of $SP_N$. This aligns with the behavior of the measure ${\cal M}$, and provides an alternative yet equivalent definition of
\begin{align}
 {\cal M}(\rho)= \inf \{  \lambda > 0 : \langle \vec{P} \rangle\in \lambda \, SP_N \} .
\label{Minkowskifunctional}
\end{align}

The Minkowski functional obeys the properties of 
\begin{enumerate}[label={(\alph*)}]
    \item Positivity: \( P(x) \in [0, \infty] \) $\forall$ \( x \in X \) and \( P(0) = 0 \).
    \item Positive Homogeneity: \( P(\lambda x) = \lambda P(x) \) $\forall$ \( \lambda \geq 0 \) and \( x \in X \).
\end{enumerate}
Additionally, if the set \( C \) is convex, the Minkowski functional \( P \) also satisfies:
\begin{enumerate}[label={(\alph*)}, resume]
    \item Subadditivity: \( P(x + y) \leq P(x) + P(y) \) $\forall$ \( x, y \in X \).
\end{enumerate}
These properties result in the Minkowski functional satisfying all the properties to be a sublinear function, which are all necessarily convex as shown below \cite{Narici2010}.

Let us consider a functional $P(tx+(1-t)y)$ and using the properties of subadditivity and 
homogenity we get: 
\begin{eqnarray}
   P(tx+(1-t)y) &\leq & P(tx) + P((1-t)y)  \nonumber \\
                & = &  t P(x) + (1-t) P(y),   \nonumber            
\end{eqnarray}
which gives us the proof of convexity
\begin{equation}
   P(tx+(1-t)y) \leq    P(x) + (1-t) P(y). 
\end{equation}
Hence the functional measure ${\cal M}(\rho) $, being a Minkowski functional, is a convex function.

\section{The magic witness $ { \cal W }(\rho) $}

In this section we show how we arrive at the form of the witness (\ref{magicmeasuresimple}).  

Consider the quantity
\begin{align}
{\cal Y }(\rho)  = \max_{  \{ \vec{a} \in \mathbb{Z}^{D^2}, a_0 = 0    \}  }  [ \frac{ \vec{a} \cdot \langle \vec{P} \rangle  -  b(\vec{a}) }{||\vec{a}  ||} ] ,
\label{magicmeasure}
\end{align}
where we include the norm $ || \vec{a} || $ such as to normalize the vector.  A similar maximization as (\ref{magicmeasurer}) is performed. For $ \vec{a} = \vec{0} $, we take the argument of the maximization to be 0, which guarantees non-negativity $ {\cal Y } (\rho) \ge 0 $. This is a necessary and sufficient criterion for magic when $ {\cal Y } (\rho) >0 $ since the numerator takes the same form as (\ref{allpolytopeboundaries}).  Using similar arguments to $ {\cal M} (\rho) $, we therefore deduce the properties: 1) Non-negativity $ {\cal Y } (\rho) \ge 0 $;
2) Invariance under Clifford unitaries $ {\cal Y } (\rho) = {\cal Y } (U_C \rho U_C^\dagger ) $; 3) Faithfulness $ {\cal Y } (\rho) = 1 $ iff $ \rho = \rho_S$, otherwise $ {\cal Y } (\rho) > 1 $.  

Using the infinity norm $ || \vec{a} ||_\infty = \max_k |a_k | $ we have 
\begin{align}
{\cal Y} (  \rho  ) =  \max_{\vec{a} \in \mathbb{Z}^{D^2}}  \left[ \frac{\vec{a} \cdot \langle \vec{P} \rangle  -  \max_{i\in[1,D_S] } \vec{a}  \cdot \vec{S}_{i}  }{ \max_k |a_k |  }  \right] .
\end{align}
The normalization factor rescales the $ \vec{a} $ such that the maximum component is $ \pm 1 $. All other entries are rational numbers in the range $ -1 \le a_k \le 1 $.   Let us call such vectors $ \vec{a}_1 $, so we have
\begin{align}
{\cal Y} (  \rho  ) =  \max_{\vec{a}_1 }  \left[ \vec{a}_1 \cdot \langle \vec{P} \rangle  -  b( \vec{a}_1)  \right] .
\label{ywita1}
\end{align}
Since any rational number is a real number, we may extend the domain of the optimization to real numbers in the range $ -1 \le a_k \le 1 $.  This is nearly (\ref{magicmeasuresimple}), but however note that there is no constraint that the largest component $ \max_k |a_k | =1  $ in (\ref{magicmeasuresimple}), which is present in (\ref{ywita1}). This constraint does not need to be enforced due to the following reasons.  
Provided the argument of (\ref{magicmeasuresimple}) is a positive number, a larger value can always be obtained by simply scaling up the vector $ \vec{a} $ until one of the components has $ |a_k | = 1 $.  Therefore, as long as the maximization is correctly performed, no constraint needs to be applied to the $ \vec{a} $, thereby arriving at  (\ref{magicmeasuresimple}).


\begin{thebibliography}{37}%
\makeatletter
\providecommand \@ifxundefined [1]{%
 \@ifx{#1\undefined}
}%
\providecommand \@ifnum [1]{%
 \ifnum #1\expandafter \@firstoftwo
 \else \expandafter \@secondoftwo
 \fi
}%
\providecommand \@ifx [1]{%
 \ifx #1\expandafter \@firstoftwo
 \else \expandafter \@secondoftwo
 \fi
}%
\providecommand \natexlab [1]{#1}%
\providecommand \enquote  [1]{``#1''}%
\providecommand \bibnamefont  [1]{#1}%
\providecommand \bibfnamefont [1]{#1}%
\providecommand \citenamefont [1]{#1}%
\providecommand \href@noop [0]{\@secondoftwo}%
\providecommand \href [0]{\begingroup \@sanitize@url \@href}%
\providecommand \@href[1]{\@@startlink{#1}\@@href}%
\providecommand \@@href[1]{\endgroup#1\@@endlink}%
\providecommand \@sanitize@url [0]{\catcode `\\12\catcode `\$12\catcode `\&12\catcode `\#12\catcode `\^12\catcode `\_12\catcode `\%12\relax}%
\providecommand \@@startlink[1]{}%
\providecommand \@@endlink[0]{}%
\providecommand \url  [0]{\begingroup\@sanitize@url \@url }%
\providecommand \@url [1]{\endgroup\@href {#1}{\urlprefix }}%
\providecommand \urlprefix  [0]{URL }%
\providecommand \Eprint [0]{\href }%
\providecommand \doibase [0]{https://doi.org/}%
\providecommand \selectlanguage [0]{\@gobble}%
\providecommand \bibinfo  [0]{\@secondoftwo}%
\providecommand \bibfield  [0]{\@secondoftwo}%
\providecommand \translation [1]{[#1]}%
\providecommand \BibitemOpen [0]{}%
\providecommand \bibitemStop [0]{}%
\providecommand \bibitemNoStop [0]{.\EOS\space}%
\providecommand \EOS [0]{\spacefactor3000\relax}%
\providecommand \BibitemShut  [1]{\csname bibitem#1\endcsname}%
\let\auto@bib@innerbib\@empty
\bibitem [{\citenamefont {Gottesman}(1998{\natexlab{a}})}]{gottesman1998heisenberg}%
  \BibitemOpen
  \bibfield  {author} {\bibinfo {author} {\bibfnamefont {D.}~\bibnamefont {Gottesman}},\ }\href@noop {} {\bibinfo {title} {The {H}eisenberg representation of quantum computers}} (\bibinfo {year} {1998}{\natexlab{a}}),\ \Eprint {https://arxiv.org/abs/arXiv:quant-ph/9807006} {arXiv:quant-ph/9807006} \BibitemShut {NoStop}%
\bibitem [{\citenamefont {Gottesman}(1998{\natexlab{b}})}]{gottesman1998talk}%
  \BibitemOpen
  \bibfield  {author} {\bibinfo {author} {\bibfnamefont {D.}~\bibnamefont {Gottesman}},\ }\href@noop {} {\bibinfo {title} {International conference on group theoretic methods in physics}} (\bibinfo {year} {1998}{\natexlab{b}}),\ \Eprint {https://arxiv.org/abs/arXiv:quant-ph/9807006} {arXiv:quant-ph/9807006} \BibitemShut {NoStop}%
\bibitem [{\citenamefont {Aaronson}\ and\ \citenamefont {Gottesman}(2004)}]{aaronson2004improved}%
  \BibitemOpen
  \bibfield  {author} {\bibinfo {author} {\bibfnamefont {S.}~\bibnamefont {Aaronson}}\ and\ \bibinfo {author} {\bibfnamefont {D.}~\bibnamefont {Gottesman}},\ }\bibfield  {title} {\bibinfo {title} {Improved simulation of stabilizer circuits},\ }\href@noop {} {\bibfield  {journal} {\bibinfo  {journal} {Physical Review A}\ }\textbf {\bibinfo {volume} {70}},\ \bibinfo {pages} {052328} (\bibinfo {year} {2004})}\BibitemShut {NoStop}%
\bibitem [{\citenamefont {Anders}\ and\ \citenamefont {Briegel}(2006)}]{anders2006fast}%
  \BibitemOpen
  \bibfield  {author} {\bibinfo {author} {\bibfnamefont {S.}~\bibnamefont {Anders}}\ and\ \bibinfo {author} {\bibfnamefont {H.~J.}\ \bibnamefont {Briegel}},\ }\bibfield  {title} {\bibinfo {title} {Fast simulation of stabilizer circuits using a graph-state representation},\ }\href@noop {} {\bibfield  {journal} {\bibinfo  {journal} {Physical Review A}\ }\textbf {\bibinfo {volume} {73}},\ \bibinfo {pages} {022334} (\bibinfo {year} {2006})}\BibitemShut {NoStop}%
\bibitem [{\citenamefont {Gottesman}(1997)}]{gottesman1997stabilizer}%
  \BibitemOpen
  \bibfield  {author} {\bibinfo {author} {\bibfnamefont {D.}~\bibnamefont {Gottesman}},\ }\href@noop {} {\emph {\bibinfo {title} {Stabilizer codes and quantum error correction}}}\ (\bibinfo  {publisher} {California Institute of Technology},\ \bibinfo {year} {1997})\BibitemShut {NoStop}%
\bibitem [{\citenamefont {Nielsen}\ and\ \citenamefont {Chuang}(2002)}]{nielsen2002quantum}%
  \BibitemOpen
  \bibfield  {author} {\bibinfo {author} {\bibfnamefont {M.~A.}\ \bibnamefont {Nielsen}}\ and\ \bibinfo {author} {\bibfnamefont {I.}~\bibnamefont {Chuang}},\ }\href@noop {} {\bibinfo {title} {Quantum computation and quantum information}} (\bibinfo {year} {2002})\BibitemShut {NoStop}%
\bibitem [{\citenamefont {Aharonov}\ \emph {et~al.}(2023)\citenamefont {Aharonov}, \citenamefont {Gao}, \citenamefont {Landau}, \citenamefont {Liu},\ and\ \citenamefont {Vazirani}}]{aharonov2023polynomial}%
  \BibitemOpen
  \bibfield  {author} {\bibinfo {author} {\bibfnamefont {D.}~\bibnamefont {Aharonov}}, \bibinfo {author} {\bibfnamefont {X.}~\bibnamefont {Gao}}, \bibinfo {author} {\bibfnamefont {Z.}~\bibnamefont {Landau}}, \bibinfo {author} {\bibfnamefont {Y.}~\bibnamefont {Liu}},\ and\ \bibinfo {author} {\bibfnamefont {U.}~\bibnamefont {Vazirani}},\ }\bibfield  {title} {\bibinfo {title} {A polynomial-time classical algorithm for noisy random circuit sampling},\ }in\ \href@noop {} {\emph {\bibinfo {booktitle} {Proceedings of the 55th Annual ACM Symposium on Theory of Computing}}}\ (\bibinfo {year} {2023})\ pp.\ \bibinfo {pages} {945--957}\BibitemShut {NoStop}%
\bibitem [{\citenamefont {Ermakov}\ \emph {et~al.}(2024)\citenamefont {Ermakov}, \citenamefont {Lychkovskiy},\ and\ \citenamefont {Byrnes}}]{ermakov2024unified}%
  \BibitemOpen
  \bibfield  {author} {\bibinfo {author} {\bibfnamefont {I.}~\bibnamefont {Ermakov}}, \bibinfo {author} {\bibfnamefont {O.}~\bibnamefont {Lychkovskiy}},\ and\ \bibinfo {author} {\bibfnamefont {T.}~\bibnamefont {Byrnes}},\ }\bibfield  {title} {\bibinfo {title} {Unified framework for efficiently computable quantum circuits},\ }\href@noop {} {\bibfield  {journal} {\bibinfo  {journal} {arXiv preprint arXiv:2401.08187}\ } (\bibinfo {year} {2024})}\BibitemShut {NoStop}%
\bibitem [{\citenamefont {Veitch}\ \emph {et~al.}(2014)\citenamefont {Veitch}, \citenamefont {Mousavian}, \citenamefont {Gottesman},\ and\ \citenamefont {Emerson}}]{veitch2014resource}%
  \BibitemOpen
  \bibfield  {author} {\bibinfo {author} {\bibfnamefont {V.}~\bibnamefont {Veitch}}, \bibinfo {author} {\bibfnamefont {S.~H.}\ \bibnamefont {Mousavian}}, \bibinfo {author} {\bibfnamefont {D.}~\bibnamefont {Gottesman}},\ and\ \bibinfo {author} {\bibfnamefont {J.}~\bibnamefont {Emerson}},\ }\bibfield  {title} {\bibinfo {title} {The resource theory of stabilizer quantum computation},\ }\href@noop {} {\bibfield  {journal} {\bibinfo  {journal} {New Journal of Physics}\ }\textbf {\bibinfo {volume} {16}},\ \bibinfo {pages} {013009} (\bibinfo {year} {2014})}\BibitemShut {NoStop}%
\bibitem [{\citenamefont {Veitch}\ \emph {et~al.}(2012{\natexlab{a}})\citenamefont {Veitch}, \citenamefont {Ferrie}, \citenamefont {Gross},\ and\ \citenamefont {Emerson}}]{veitch2012negative}%
  \BibitemOpen
  \bibfield  {author} {\bibinfo {author} {\bibfnamefont {V.}~\bibnamefont {Veitch}}, \bibinfo {author} {\bibfnamefont {C.}~\bibnamefont {Ferrie}}, \bibinfo {author} {\bibfnamefont {D.}~\bibnamefont {Gross}},\ and\ \bibinfo {author} {\bibfnamefont {J.}~\bibnamefont {Emerson}},\ }\bibfield  {title} {\bibinfo {title} {Negative quasi-probability as a resource for quantum computation},\ }\href@noop {} {\bibfield  {journal} {\bibinfo  {journal} {New Journal of Physics}\ }\textbf {\bibinfo {volume} {14}},\ \bibinfo {pages} {113011} (\bibinfo {year} {2012}{\natexlab{a}})}\BibitemShut {NoStop}%
\bibitem [{\citenamefont {Mari}\ and\ \citenamefont {Eisert}(2012)}]{mari2012positive}%
  \BibitemOpen
  \bibfield  {author} {\bibinfo {author} {\bibfnamefont {A.}~\bibnamefont {Mari}}\ and\ \bibinfo {author} {\bibfnamefont {J.}~\bibnamefont {Eisert}},\ }\bibfield  {title} {\bibinfo {title} {Positive wigner functions render classical simulation of quantum computation efficient},\ }\href@noop {} {\bibfield  {journal} {\bibinfo  {journal} {Physical review letters}\ }\textbf {\bibinfo {volume} {109}},\ \bibinfo {pages} {230503} (\bibinfo {year} {2012})}\BibitemShut {NoStop}%
\bibitem [{\citenamefont {Howard}\ and\ \citenamefont {Campbell}(2017)}]{howard2017application}%
  \BibitemOpen
  \bibfield  {author} {\bibinfo {author} {\bibfnamefont {M.}~\bibnamefont {Howard}}\ and\ \bibinfo {author} {\bibfnamefont {E.}~\bibnamefont {Campbell}},\ }\bibfield  {title} {\bibinfo {title} {Application of a resource theory for magic states to fault-tolerant quantum computing},\ }\href@noop {} {\bibfield  {journal} {\bibinfo  {journal} {Physical review letters}\ }\textbf {\bibinfo {volume} {118}},\ \bibinfo {pages} {090501} (\bibinfo {year} {2017})}\BibitemShut {NoStop}%
\bibitem [{\citenamefont {Vidal}\ and\ \citenamefont {Tarrach}(1999)}]{vidal1999robustness}%
  \BibitemOpen
  \bibfield  {author} {\bibinfo {author} {\bibfnamefont {G.}~\bibnamefont {Vidal}}\ and\ \bibinfo {author} {\bibfnamefont {R.}~\bibnamefont {Tarrach}},\ }\bibfield  {title} {\bibinfo {title} {Robustness of entanglement},\ }\href {https://doi.org/10.1103/physreva.59.141} {\bibfield  {journal} {\bibinfo  {journal} {Physical Review A}\ }\textbf {\bibinfo {volume} {59}},\ \bibinfo {pages} {141–155} (\bibinfo {year} {1999})}\BibitemShut {NoStop}%
\bibitem [{\citenamefont {Seddon}\ \emph {et~al.}(2021)\citenamefont {Seddon}, \citenamefont {Regula}, \citenamefont {Pashayan}, \citenamefont {Ouyang},\ and\ \citenamefont {Campbell}}]{seddon2021quantifying}%
  \BibitemOpen
  \bibfield  {author} {\bibinfo {author} {\bibfnamefont {J.~R.}\ \bibnamefont {Seddon}}, \bibinfo {author} {\bibfnamefont {B.}~\bibnamefont {Regula}}, \bibinfo {author} {\bibfnamefont {H.}~\bibnamefont {Pashayan}}, \bibinfo {author} {\bibfnamefont {Y.}~\bibnamefont {Ouyang}},\ and\ \bibinfo {author} {\bibfnamefont {E.~T.}\ \bibnamefont {Campbell}},\ }\bibfield  {title} {\bibinfo {title} {Quantifying quantum speedups: Improved classical simulation from tighter magic monotones},\ }\href@noop {} {\bibfield  {journal} {\bibinfo  {journal} {PRX Quantum}\ }\textbf {\bibinfo {volume} {2}},\ \bibinfo {pages} {010345} (\bibinfo {year} {2021})}\BibitemShut {NoStop}%
\bibitem [{\citenamefont {Bravyi}\ and\ \citenamefont {Gosset}(2016)}]{bravyi2016improved}%
  \BibitemOpen
  \bibfield  {author} {\bibinfo {author} {\bibfnamefont {S.}~\bibnamefont {Bravyi}}\ and\ \bibinfo {author} {\bibfnamefont {D.}~\bibnamefont {Gosset}},\ }\bibfield  {title} {\bibinfo {title} {Improved classical simulation of quantum circuits dominated by {C}lifford gates},\ }\href@noop {} {\bibfield  {journal} {\bibinfo  {journal} {Physical review letters}\ }\textbf {\bibinfo {volume} {116}},\ \bibinfo {pages} {250501} (\bibinfo {year} {2016})}\BibitemShut {NoStop}%
\bibitem [{\citenamefont {Bravyi}\ \emph {et~al.}(2016)\citenamefont {Bravyi}, \citenamefont {Smith},\ and\ \citenamefont {Smolin}}]{bravyi2016trading}%
  \BibitemOpen
  \bibfield  {author} {\bibinfo {author} {\bibfnamefont {S.}~\bibnamefont {Bravyi}}, \bibinfo {author} {\bibfnamefont {G.}~\bibnamefont {Smith}},\ and\ \bibinfo {author} {\bibfnamefont {J.~A.}\ \bibnamefont {Smolin}},\ }\bibfield  {title} {\bibinfo {title} {Trading classical and quantum computational resources},\ }\href@noop {} {\bibfield  {journal} {\bibinfo  {journal} {Physical Review X}\ }\textbf {\bibinfo {volume} {6}},\ \bibinfo {pages} {021043} (\bibinfo {year} {2016})}\BibitemShut {NoStop}%
\bibitem [{\citenamefont {Bravyi}\ \emph {et~al.}(2019)\citenamefont {Bravyi}, \citenamefont {Browne}, \citenamefont {Calpin}, \citenamefont {Campbell}, \citenamefont {Gosset},\ and\ \citenamefont {Howard}}]{bravyi2019simulation}%
  \BibitemOpen
  \bibfield  {author} {\bibinfo {author} {\bibfnamefont {S.}~\bibnamefont {Bravyi}}, \bibinfo {author} {\bibfnamefont {D.}~\bibnamefont {Browne}}, \bibinfo {author} {\bibfnamefont {P.}~\bibnamefont {Calpin}}, \bibinfo {author} {\bibfnamefont {E.}~\bibnamefont {Campbell}}, \bibinfo {author} {\bibfnamefont {D.}~\bibnamefont {Gosset}},\ and\ \bibinfo {author} {\bibfnamefont {M.}~\bibnamefont {Howard}},\ }\bibfield  {title} {\bibinfo {title} {Simulation of quantum circuits by low-rank stabilizer decompositions},\ }\href@noop {} {\bibfield  {journal} {\bibinfo  {journal} {Quantum}\ }\textbf {\bibinfo {volume} {3}},\ \bibinfo {pages} {181} (\bibinfo {year} {2019})}\BibitemShut {NoStop}%
\bibitem [{\citenamefont {Beverland}\ \emph {et~al.}(2020)\citenamefont {Beverland}, \citenamefont {Campbell}, \citenamefont {Howard},\ and\ \citenamefont {Kliuchnikov}}]{beverland2020lower}%
  \BibitemOpen
  \bibfield  {author} {\bibinfo {author} {\bibfnamefont {M.}~\bibnamefont {Beverland}}, \bibinfo {author} {\bibfnamefont {E.}~\bibnamefont {Campbell}}, \bibinfo {author} {\bibfnamefont {M.}~\bibnamefont {Howard}},\ and\ \bibinfo {author} {\bibfnamefont {V.}~\bibnamefont {Kliuchnikov}},\ }\bibfield  {title} {\bibinfo {title} {Lower bounds on the non-{C}lifford resources for quantum computations},\ }\href@noop {} {\bibfield  {journal} {\bibinfo  {journal} {Quantum Science and Technology}\ }\textbf {\bibinfo {volume} {5}},\ \bibinfo {pages} {035009} (\bibinfo {year} {2020})}\BibitemShut {NoStop}%
\bibitem [{\citenamefont {Leone}\ \emph {et~al.}(2022)\citenamefont {Leone}, \citenamefont {Oliviero},\ and\ \citenamefont {Hamma}}]{leone2022renyientropy}%
  \BibitemOpen
  \bibfield  {author} {\bibinfo {author} {\bibfnamefont {L.}~\bibnamefont {Leone}}, \bibinfo {author} {\bibfnamefont {S.~F.~E.}\ \bibnamefont {Oliviero}},\ and\ \bibinfo {author} {\bibfnamefont {A.}~\bibnamefont {Hamma}},\ }\bibfield  {title} {\bibinfo {title} {Stabilizer r\'enyi entropy},\ }\href {https://doi.org/10.1103/PhysRevLett.128.050402} {\bibfield  {journal} {\bibinfo  {journal} {Phys. Rev. Lett.}\ }\textbf {\bibinfo {volume} {128}},\ \bibinfo {pages} {050402} (\bibinfo {year} {2022})}\BibitemShut {NoStop}%
\bibitem [{\citenamefont {Leone}\ and\ \citenamefont {Bittel}(2024)}]{leone2024stabilizerentropies}%
  \BibitemOpen
  \bibfield  {author} {\bibinfo {author} {\bibfnamefont {L.}~\bibnamefont {Leone}}\ and\ \bibinfo {author} {\bibfnamefont {L.}~\bibnamefont {Bittel}},\ }\href {https://arxiv.org/abs/2404.11652} {\bibinfo {title} {Stabilizer entropies are monotones for magic-state resource theory}} (\bibinfo {year} {2024}),\ \Eprint {https://arxiv.org/abs/quant-ph/2404.11652} {arXiv:quant-ph/2404.11652} \BibitemShut {NoStop}%
\bibitem [{\citenamefont {Haug}\ and\ \citenamefont {Piroli}(2023)}]{haug2023stabilizerentropies}%
  \BibitemOpen
  \bibfield  {author} {\bibinfo {author} {\bibfnamefont {T.}~\bibnamefont {Haug}}\ and\ \bibinfo {author} {\bibfnamefont {L.}~\bibnamefont {Piroli}},\ }\bibfield  {title} {\bibinfo {title} {Stabilizer entropies and nonstabilizerness monotones},\ }\href {https://doi.org/10.22331/q-2023-08-28-1092} {\bibfield  {journal} {\bibinfo  {journal} {Quantum}\ }\textbf {\bibinfo {volume} {7}},\ \bibinfo {pages} {1092} (\bibinfo {year} {2023})}\BibitemShut {NoStop}%
\bibitem [{\citenamefont {Oliviero}\ \emph {et~al.}(2022)\citenamefont {Oliviero}, \citenamefont {Leone}, \citenamefont {Hamma},\ and\ \citenamefont {Lloyd}}]{oliviero2022measuring}%
  \BibitemOpen
  \bibfield  {author} {\bibinfo {author} {\bibfnamefont {S.~F.}\ \bibnamefont {Oliviero}}, \bibinfo {author} {\bibfnamefont {L.}~\bibnamefont {Leone}}, \bibinfo {author} {\bibfnamefont {A.}~\bibnamefont {Hamma}},\ and\ \bibinfo {author} {\bibfnamefont {S.}~\bibnamefont {Lloyd}},\ }\bibfield  {title} {\bibinfo {title} {Measuring magic on a quantum processor},\ }\href@noop {} {\bibfield  {journal} {\bibinfo  {journal} {npj Quantum Information}\ }\textbf {\bibinfo {volume} {8}},\ \bibinfo {pages} {148} (\bibinfo {year} {2022})}\BibitemShut {NoStop}%
\bibitem [{\citenamefont {Hahn}\ \emph {et~al.}(2022)\citenamefont {Hahn}, \citenamefont {Ferraro}, \citenamefont {Hultquist}, \citenamefont {Ferrini},\ and\ \citenamefont {Garc{\'\i}a-{\'A}lvarez}}]{hahn2022quantifying}%
  \BibitemOpen
  \bibfield  {author} {\bibinfo {author} {\bibfnamefont {O.}~\bibnamefont {Hahn}}, \bibinfo {author} {\bibfnamefont {A.}~\bibnamefont {Ferraro}}, \bibinfo {author} {\bibfnamefont {L.}~\bibnamefont {Hultquist}}, \bibinfo {author} {\bibfnamefont {G.}~\bibnamefont {Ferrini}},\ and\ \bibinfo {author} {\bibfnamefont {L.}~\bibnamefont {Garc{\'\i}a-{\'A}lvarez}},\ }\bibfield  {title} {\bibinfo {title} {Quantifying qubit magic resource with {G}ottesman-{K}itaev-{P}reskill encoding},\ }\href@noop {} {\bibfield  {journal} {\bibinfo  {journal} {Physical Review Letters}\ }\textbf {\bibinfo {volume} {128}},\ \bibinfo {pages} {210502} (\bibinfo {year} {2022})}\BibitemShut {NoStop}%
\bibitem [{\citenamefont {Haug}\ and\ \citenamefont {Kim}(2023)}]{haug2023scalablemeasures}%
  \BibitemOpen
  \bibfield  {author} {\bibinfo {author} {\bibfnamefont {T.}~\bibnamefont {Haug}}\ and\ \bibinfo {author} {\bibfnamefont {M.}~\bibnamefont {Kim}},\ }\bibfield  {title} {\bibinfo {title} {Scalable measures of magic resource for quantum computers},\ }\href {https://doi.org/10.1103/PRXQuantum.4.010301} {\bibfield  {journal} {\bibinfo  {journal} {PRX Quantum}\ }\textbf {\bibinfo {volume} {4}},\ \bibinfo {pages} {010301} (\bibinfo {year} {2023})}\BibitemShut {NoStop}%
\bibitem [{\citenamefont {Veitch}\ \emph {et~al.}(2012{\natexlab{b}})\citenamefont {Veitch}, \citenamefont {Ferrie}, \citenamefont {Gross},\ and\ \citenamefont {Emerson}}]{veitch2012negativequasiprobability}%
  \BibitemOpen
  \bibfield  {author} {\bibinfo {author} {\bibfnamefont {V.}~\bibnamefont {Veitch}}, \bibinfo {author} {\bibfnamefont {C.}~\bibnamefont {Ferrie}}, \bibinfo {author} {\bibfnamefont {D.}~\bibnamefont {Gross}},\ and\ \bibinfo {author} {\bibfnamefont {J.}~\bibnamefont {Emerson}},\ }\bibfield  {title} {\bibinfo {title} {Negative quasi-probability as a resource for quantum computation},\ }\href {https://doi.org/10.1088/1367-2630/14/11/113011} {\bibfield  {journal} {\bibinfo  {journal} {New Journal of Physics}\ }\textbf {\bibinfo {volume} {14}},\ \bibinfo {pages} {113011} (\bibinfo {year} {2012}{\natexlab{b}})}\BibitemShut {NoStop}%
\bibitem [{\citenamefont {Wang}\ \emph {et~al.}(2020)\citenamefont {Wang}, \citenamefont {Wilde},\ and\ \citenamefont {Su}}]{wang2020thauma}%
  \BibitemOpen
  \bibfield  {author} {\bibinfo {author} {\bibfnamefont {X.}~\bibnamefont {Wang}}, \bibinfo {author} {\bibfnamefont {M.~M.}\ \bibnamefont {Wilde}},\ and\ \bibinfo {author} {\bibfnamefont {Y.}~\bibnamefont {Su}},\ }\bibfield  {title} {\bibinfo {title} {Efficiently computable bounds for magic state distillation},\ }\href {https://doi.org/10.1103/PhysRevLett.124.090505} {\bibfield  {journal} {\bibinfo  {journal} {Phys. Rev. Lett.}\ }\textbf {\bibinfo {volume} {124}},\ \bibinfo {pages} {090505} (\bibinfo {year} {2020})}\BibitemShut {NoStop}%
\bibitem [{\citenamefont {Wang}\ \emph {et~al.}(2019)\citenamefont {Wang}, \citenamefont {Wilde},\ and\ \citenamefont {Su}}]{wang2019quantifyingmagic}%
  \BibitemOpen
  \bibfield  {author} {\bibinfo {author} {\bibfnamefont {X.}~\bibnamefont {Wang}}, \bibinfo {author} {\bibfnamefont {M.~M.}\ \bibnamefont {Wilde}},\ and\ \bibinfo {author} {\bibfnamefont {Y.}~\bibnamefont {Su}},\ }\bibfield  {title} {\bibinfo {title} {Quantifying the magic of quantum channels},\ }\href {https://doi.org/10.1088/1367-2630/ab451d} {\bibfield  {journal} {\bibinfo  {journal} {New Journal of Physics}\ }\textbf {\bibinfo {volume} {21}},\ \bibinfo {pages} {103002} (\bibinfo {year} {2019})}\BibitemShut {NoStop}%
\bibitem [{\citenamefont {Koukoulekidis}\ and\ \citenamefont {Jennings}(2022)}]{koukoulekidis2022wignernegativity}%
  \BibitemOpen
  \bibfield  {author} {\bibinfo {author} {\bibfnamefont {N.}~\bibnamefont {Koukoulekidis}}\ and\ \bibinfo {author} {\bibfnamefont {D.}~\bibnamefont {Jennings}},\ }\bibfield  {title} {\bibinfo {title} {Constraints on magic state protocols from the statistical mechanics of {W}igner negativity},\ }\href@noop {} {\bibfield  {journal} {\bibinfo  {journal} {npj Quantum Information}\ }\textbf {\bibinfo {volume} {8}} (\bibinfo {year} {2022})}\BibitemShut {NoStop}%
\bibitem [{\citenamefont {Pashayan}\ \emph {et~al.}(2015)\citenamefont {Pashayan}, \citenamefont {Wallman},\ and\ \citenamefont {Bartlett}}]{pashayan2015quasiprobabilities}%
  \BibitemOpen
  \bibfield  {author} {\bibinfo {author} {\bibfnamefont {H.}~\bibnamefont {Pashayan}}, \bibinfo {author} {\bibfnamefont {J.~J.}\ \bibnamefont {Wallman}},\ and\ \bibinfo {author} {\bibfnamefont {S.~D.}\ \bibnamefont {Bartlett}},\ }\bibfield  {title} {\bibinfo {title} {Estimating outcome probabilities of quantum circuits using quasiprobabilities},\ }\href {https://doi.org/10.1103/PhysRevLett.115.070501} {\bibfield  {journal} {\bibinfo  {journal} {Phys. Rev. Lett.}\ }\textbf {\bibinfo {volume} {115}},\ \bibinfo {pages} {070501} (\bibinfo {year} {2015})}\BibitemShut {NoStop}%
\bibitem [{\citenamefont {Kulikov}\ \emph {et~al.}(2024)\citenamefont {Kulikov}, \citenamefont {Yashin}, \citenamefont {Fedorov},\ and\ \citenamefont {Kiktenko}}]{kulikov2024minimizingnegativity}%
  \BibitemOpen
  \bibfield  {author} {\bibinfo {author} {\bibfnamefont {D.~A.}\ \bibnamefont {Kulikov}}, \bibinfo {author} {\bibfnamefont {V.~I.}\ \bibnamefont {Yashin}}, \bibinfo {author} {\bibfnamefont {A.~K.}\ \bibnamefont {Fedorov}},\ and\ \bibinfo {author} {\bibfnamefont {E.~O.}\ \bibnamefont {Kiktenko}},\ }\bibfield  {title} {\bibinfo {title} {Minimizing the negativity of quantum circuits in overcomplete quasiprobability representations},\ }\href {https://doi.org/10.1103/PhysRevA.109.012219} {\bibfield  {journal} {\bibinfo  {journal} {Phys. Rev. A}\ }\textbf {\bibinfo {volume} {109}},\ \bibinfo {pages} {012219} (\bibinfo {year} {2024})}\BibitemShut {NoStop}%
\bibitem [{\citenamefont {Campbell}(2011)}]{campbell2011catalysis}%
  \BibitemOpen
  \bibfield  {author} {\bibinfo {author} {\bibfnamefont {E.~T.}\ \bibnamefont {Campbell}},\ }\bibfield  {title} {\bibinfo {title} {Catalysis and activation of magic states in fault-tolerant architectures},\ }\href@noop {} {\bibfield  {journal} {\bibinfo  {journal} {Physical Review A}\ }\textbf {\bibinfo {volume} {83}},\ \bibinfo {pages} {032317} (\bibinfo {year} {2011})}\BibitemShut {NoStop}%
\bibitem [{\citenamefont {Gross}(2006)}]{gross2006hudson}%
  \BibitemOpen
  \bibfield  {author} {\bibinfo {author} {\bibfnamefont {D.}~\bibnamefont {Gross}},\ }\bibfield  {title} {\bibinfo {title} {Hudson’s theorem for finite-dimensional quantum systems},\ }\href@noop {} {\bibfield  {journal} {\bibinfo  {journal} {Journal of mathematical physics}\ }\textbf {\bibinfo {volume} {47}} (\bibinfo {year} {2006})}\BibitemShut {NoStop}%
\bibitem [{\citenamefont {Gr{\"u}nbaum}\ \emph {et~al.}(1967)\citenamefont {Gr{\"u}nbaum}, \citenamefont {Klee}, \citenamefont {Perles},\ and\ \citenamefont {Shephard}}]{grunbaum1967convex}%
  \BibitemOpen
  \bibfield  {author} {\bibinfo {author} {\bibfnamefont {B.}~\bibnamefont {Gr{\"u}nbaum}}, \bibinfo {author} {\bibfnamefont {V.}~\bibnamefont {Klee}}, \bibinfo {author} {\bibfnamefont {M.~A.}\ \bibnamefont {Perles}},\ and\ \bibinfo {author} {\bibfnamefont {G.~C.}\ \bibnamefont {Shephard}},\ }\href@noop {} {\emph {\bibinfo {title} {Convex polytopes}}},\ Vol.~\bibinfo {volume} {16}\ (\bibinfo  {publisher} {Springer},\ \bibinfo {year} {1967})\BibitemShut {NoStop}%
\bibitem [{\citenamefont {Heinrich}\ and\ \citenamefont {Gross}(2019)}]{heinrich2019robustness}%
  \BibitemOpen
  \bibfield  {author} {\bibinfo {author} {\bibfnamefont {M.}~\bibnamefont {Heinrich}}\ and\ \bibinfo {author} {\bibfnamefont {D.}~\bibnamefont {Gross}},\ }\bibfield  {title} {\bibinfo {title} {Robustness of magic and symmetries of the stabiliser polytope},\ }\href@noop {} {\bibfield  {journal} {\bibinfo  {journal} {Quantum}\ }\textbf {\bibinfo {volume} {3}},\ \bibinfo {pages} {132} (\bibinfo {year} {2019})}\BibitemShut {NoStop}%
\bibitem [{\citenamefont {Reichardt}(2009)}]{reichardt2009quantum}%
  \BibitemOpen
  \bibfield  {author} {\bibinfo {author} {\bibfnamefont {B.~W.}\ \bibnamefont {Reichardt}},\ }\bibfield  {title} {\bibinfo {title} {Quantum universality by state distillation},\ }\href@noop {} {\bibfield  {journal} {\bibinfo  {journal} {Quantum Info. Comput.}\ }\textbf {\bibinfo {volume} {9}},\ \bibinfo {pages} {1030} (\bibinfo {year} {2009})}\BibitemShut {NoStop}%
\bibitem [{\citenamefont {Streltsov}\ \emph {et~al.}(2017)\citenamefont {Streltsov}, \citenamefont {Adesso},\ and\ \citenamefont {Plenio}}]{streltsov2017colloquium}%
  \BibitemOpen
  \bibfield  {author} {\bibinfo {author} {\bibfnamefont {A.}~\bibnamefont {Streltsov}}, \bibinfo {author} {\bibfnamefont {G.}~\bibnamefont {Adesso}},\ and\ \bibinfo {author} {\bibfnamefont {M.~B.}\ \bibnamefont {Plenio}},\ }\bibfield  {title} {\bibinfo {title} {Colloquium: Quantum coherence as a resource},\ }\href@noop {} {\bibfield  {journal} {\bibinfo  {journal} {Reviews of Modern Physics}\ }\textbf {\bibinfo {volume} {89}},\ \bibinfo {pages} {041003} (\bibinfo {year} {2017})}\BibitemShut {NoStop}%
\bibitem [{\citenamefont {Narici}\ and\ \citenamefont {Beckenstein}(2010)}]{Narici2010}%
  \BibitemOpen
  \bibfield  {author} {\bibinfo {author} {\bibfnamefont {L.}~\bibnamefont {Narici}}\ and\ \bibinfo {author} {\bibfnamefont {E.}~\bibnamefont {Beckenstein}},\ }\href {https://doi.org/10.1201/9781584888673} {\emph {\bibinfo {title} {Topological Vector Spaces}}},\ \bibinfo {edition} {2nd}\ ed.\ (\bibinfo  {publisher} {Chapman and Hall/CRC},\ \bibinfo {year} {2010})\ \bibinfo {note} {pp. 192-193}\BibitemShut {NoStop}%
\end{thebibliography}
\end{document}